\documentclass[prd,twocolumn,nofootinbib,showpacs,superscriptaddress]{revtex4}

\usepackage{aas_macros}
\usepackage{amsfonts}
\usepackage{amsmath}
\usepackage{bm}
\usepackage{graphicx}
\usepackage[latin1]{inputenc}
\usepackage{latexsym}
\usepackage{rotating}

\newcommand{\mb}[1]{\mbox{\boldmath $#1$}}

\def \met  {\mbox{g}}

\def \sch  {\mbox{\tiny Sch}}
\newcommand{\ZM}{{\mbox{\tiny ZM}}}
\newcommand{\RW}{{\mbox{\tiny RW}}}
\newcommand{\CPM}{{\mbox{\tiny CPM}}}

\begin{document}

\title[A Finite Element Computation of the Gravitational Radiation emitted by
a Point-like object orbiting a Non-rotating Black Hole]
{A Finite Element Computation of the Gravitational Radiation emitted by
a Point-like object orbiting a Non-rotating Black Hole}

\author{Carlos F. Sopuerta}
\affiliation{Institute for Gravitational Physics and Geometry and
Center for Gravitational Wave Physics, \\
Department of Astronomy \& Astrophysics, \\
The Pennsylvania State University, University Park, PA 16802, USA}
\author{Pablo Laguna}
\affiliation{Institute for Gravitational Physics and Geometry and
Center for Gravitational Wave Physics, \\
Department of Astronomy \& Astrophysics, \\
The Pennsylvania State University, University Park, PA 16802, USA}
\affiliation{Department of Physics, The Pennsylvania State University,
University Park, PA 16802, USA}

\date{\today}

\begin{abstract}
The description of extreme-mass-ratio binary systems in the inspiral phase is a 
challenging problem in gravitational wave physics with significant relevance for 
the space interferometer LISA.   The main difficulty lies in the evaluation of the 
effects of the small body's gravitational field on itself.  To that end, an accurate 
computation of the perturbations produced by the small body with respect the 
background geometry of the large object, a massive black hole, is required.
In this paper we present a new computational approach based on Finite Element 
Methods to solve the master equations describing perturbations of non-rotating 
black holes due to an orbiting point-like object.   The numerical computations
are carried out in the time domain by using evolution algorithms for wave-type 
equations.   We show the accuracy of the method by comparing our calculations
with previous results in the literature.  Finally, we discuss the relevance of
this method for achieving accurate descriptions of extreme-mass-ratio binaries.
\end{abstract}

\pacs{04.30.Db, 04.40.Dg, 95.30.Sf, 97.10.Sj}

\maketitle

\section{Introduction}\label{intro}
Extreme-Mass-Ratio Binaries (EMRBs) in the inspiral stage of their evolution
are considered to be a primary source of gravitational radiation~\cite{Barack:2003fp,Gair:2004ea}
to be detected by the proposed laser interferometric space antenna
LISA~\cite{Vitale:2002qv,Danzmann:2003ad,Danzmann:2003tv,Prince:2003aa}.
They consist of a ``small'' object, such a main sequence star, a stellar mass
black hole, or a neutron star, with mass $m$ ranging from $1 M_\odot$ to $10^2 M_\odot$,
orbiting a massive black hole (MBH) with mass $M$ ranging from $10^3 M_\odot$
(if we consider the case of intermediate mass black holes in globular clusters)
to $10^9 M_\odot$
(the case of big supermassive black holes sitting in the center of galaxies).
This translates to EMRBs with mass ratios, $\mu = m/M\,,$ in the range
$10^{-1} - 10^{-9}\,$.
In order to exploit this type of systems through LISA, it is crucial to have a
good theoretical understanding of their evolution, good enough
to produce accurate waveform templates in support of data analysis efforts.
Because there is no significant coupling between the strong curvature effects
produced by the MBH and its companion, relativistic perturbation theory is a well
suited tool to study EMBRs.   Clearly, the accuracy of this approximation depends on the
smallness of the mass ratio $\mu$.

The challenge in modeling EMRBs is to compute the perturbations generated by the small 
body in the (background) gravitational field of the MBH, and how these perturbations
affect the motion of the small body itself.   This problem has been known in
literature as the {\em radiation reaction} problem.  This is an old problem and
several approaches to deal with it have been proposed (see the recent review
by Poisson~\cite{Poisson:2004lr} and the contributions to~\cite{Lousto:2005cq}).
The most extended approach consists in modelling the small object by using a
point-like description and then, to describe the {\em radiation reaction}
effects on the dynamics as the action of a local {\em self-force} that is
responsible for the deviations from geodesic motion.  A consistent derivation
of the equations of motion coming out from this set up was given by Mino,
Sasaki and Tanaka~\cite{Mino:1997nk}, and later, adopting an axiomatic
approach, by Quinn and Wald~\cite{Quinn:1997am} (see also~\cite{Detweiler:2002mi}).
However, these works only provide a formal prescription for the description
of the orbital motion.  For the practical calculations of the self-force
some techniques have been proposed: the {\em mode-sum}
scheme~\cite{Barack:1999wf,Barack:2000eh,Barack:2001bw,Barack:2001gx}, and a
regularization scheme based on zeta-function regularization
techniques~\cite{Lousto:1999za} (see~\cite{Lousto:2005co} for a recent progress report).

The computation of the self-force and waveforms, and any other physical
relevant information related to the inspiral due to radiation reaction constitute
the main challenge of this problem.  One possible way is to resort to analytic
techniques by adding extra approximations to the problem, similar to those from
post-Newtonian methods.   However, the results may not be applicable to
situations of physical relevance involving highly spinning MBHs and very
eccentric orbits.  To make computations without making further simplifications of the
problem, numerical techniques appear to be a necessary tool.
It is important to distinguish between frequency-domain and time-domain calculations.
The frequency domain approach has been used for a long time; it provides accurate
results for the computation of quasinormal modes and 
frequencies~\cite{Vishveshwara:1970vi,Chandrasekhar:1975qn}.
However, the frequency-domain approach has more difficulties when we are interested
in computing the waves originated from highly eccentric orbits since one has to
sum over a large number of modes to obtain a good accuracy.  In this sense,
calculations in the time-domain can be more efficient for obtaining accurate
waveforms for the physical situations of relevance.

However, the time-domain numerical approach has to face a challenge, which consists of
dealing with the different physical scales (both spatial and temporal) present
in the problem and that expand over several orders of magnitude.
Specifically, one needs to handle not only large
wavelength scales comparable to the massive black hole, but also to resolve the
scales in the vicinity of the small object where radiation reaction effects
play a crucial role.    The conclusion is that we need to incorporate adaptive
schemes in our numerical algorithms in order to provide the resolution that
every region in the physical domain requires.  Since the small object is going
to be moving through the domain (unless we choose a very particular coordinate
system), it is convenient to allow the adaptive scheme to change in time to distribute
properly the resolution.  Our choice to deal with these issues is the Finite Element
Method (FEM), which is a numerical technique where adaptivity can be implemented
in a natural way.   The FEM has other properties, which we will discuss in this
paper, that make it very suitable to be used for the description of EMRBs and also
for other physical systems that are the subject of investigation in
Numerical Relativity.   In a recent work~\cite{Sopuerta:2005rd}, we have already
tested Adaptive Mesh Refinement techniques intrinsic to the FEM in a toy model 
consisting of a particle orbiting a black hole in the context of scalar gravity, 
where we have shown how an adaptive scheme can provide better accuracy than a 
non-adaptive scheme with an equivalent computational cost.

In this paper we use the FEM to perform time-domain simulations of a 
point-like object orbiting in geodesics (no radiation-reaction) around a 
non-rotating MBH, and compute physically relevant quantities like energy and angular
momentum emitted in gravitational waves and waveforms.  This type of calculations
constitute a good touchstone to evaluate the Finite Element (FE) techniques that 
we present in this paper, specially in relation to use this type of computations 
for the evaluation of the self-force on the particle.
Since the MBH is non-rotating MBH the problem can be reduced to solve the
one-dimensional Partial Differential Equations (PDEs)
of black hole perturbations theory.  These equations, in the Regge-Wheeler
gauge, reduce to a master equation from which the metric perturbations can be
fully recovered.   The master equation for axial modes is known as the Regge-Wheeler
equation, and for polar modes as the Zerilli-Moncrief equation.
In this paper, instead of using the Regge-Wheeler function, we use a modification
originally proposed by Cunningham, Price and Moncrief~\cite{Cunningham:1978cp} that puts
the axial modes on an equal footing with polar modes, as described by the Zerilli-Moncrief
function, in relation to computing energy and angular momentum luminosities, and waveforms.

The plan of the paper is the following: In section~\ref{bhpert} we summarize the
main results from (non-rotating) black-hole perturbation theory that we need in 
our computations, including the explicit form of the source terms coming from
the particle energy-momentum tensor.  
As far as we know, the expressions we present here for the sources associated with
the axial modes, described by the Cunningham-Price-Moncrief master function, are new.
We also perform an analysis of the discontinuities 
in the master equations due to the Dirac delta distributions that the 
source terms exhibit.
In section~\ref{femformulation} we describe the numerical 
framework.  We use a FE discretization for the spatial domain and a Finite 
Difference discretization in time.
We start with the discretization of the domain, consisting of dividing the 
computational domain into disjoint subdomains (the {\em elements}). Then, we 
describe the FE functional spaces, which are finite-dimensional functional 
spaces used to approximate locally (at each element) our solution.  The next 
step is the derivation of the {\em weak} form of the master equations, which 
is an integral form.  It is important to remark that FE algorithms are derived 
from the integral form of the equations, in contrast with other numerical 
techniques where the differential form is used to obtain a discretization.
From the weak form, we obtain the spatial discretization by imposing the 
vanishing of the residuals of our equations, which basically means to impose 
the vanishing of the components of the equations with respect to a basis of 
functions constructed from the FE functional spaces.  This process leads to a 
coupled system of Ordinary Differential Equations (ODEs) which has a close 
analogy with the equations governing the behaviour of a system of coupled 
oscillators.  A very important point in the discretization process is the fact 
that, because the FE formulation is based on an integral form of the equations, 
we obtain automatically a discretization of the sources containing Dirac delta 
distributions and its first derivative without having to resort to sequences 
of functions approaching in some limit the Dirac delta, we just use the
properties of these distributions at the analytic level in the {\em weak} form 
of the equations.  To solve the resulting ODEs we introduce a collection of 
evolution algorithms to solve the equations in second-order form and which 
have parameters that allow us to control the appearance of spurious 
high-frequency modes, which are common in systems like the one we are studying 
having a very localized source.  We finish this section by discussing the 
structure of the mesh, in particular how adaptivity is implemented and how we 
can change in time this structure as the particle moves.   In 
section~\ref{results} we discuss the performance of the FE numerical code we 
have developed and compare results regarding the computations of energy and
angular momentum radiated with previous works in the literature, showing in this
way the accuracy that this method is able to achieve.  We conclude in
section~\ref{discussion}, where we discuss the convenience of using the FEM for
the simulations of EMRBs in the light of the results of this paper and describe
possible ways to proceed in the future to make this goal a reality.  Finally, we have
included two appendices:  In Appendix~\ref{geodesic}, we summarize the geodesic 
equations of motion for the particle, and in Appendix~\ref{GaussLegendre}, we
briefly describe the Gauss-Legendre quadrature method for evaluating numerically 
some of the integrals that appear in the FE discretization of our equations.

The conventions that we follow throughout this work are:
Greek letters are used to denote spacetime indices;
capital Latin letters are used for indices in the time-radial
part of the metric;  lower-case Latin indices are used for
 the spherical sector of the metric.  We use physical
units in which $G = c = 1$.

\section{Summary of Perturbation Theory for non-rotating Black Holes\label{bhpert}}
Perturbation theory of black holes has a long history. It goes back to the 
seminal work by Regge and Wheeler~\cite{Regge:1957rw} and later by 
Vishveshwara, Zerilli and 
Moncrief~\cite{Vishveshwara:1970vi,Zerilli:1970fj,Moncrief:1974vm} for 
non-rotating black holes and to
Teukolsky~\cite{Teukolsky:1972le,Teukolsky:1973ap} for rotating ones.  At present,
in the case of non-rotating black holes the metric perturbative scheme is
completely developed and well understood at the linear level
(see~\cite{Chandrasekhar:1992bo,Kokkotas:1999lr,Nollert:1999re,Sarbach:2001qq,Martel:2005ir,Martel:2003th,Nagar:2005ea}
for reviews).  In the particular case of perturbations induced by an orbiting
point-like object, which is the situation we are interested in this paper, there are a
number of works on it~\cite{Zerilli:1970la,Davis:1972pa,Lousto:1997wf,Martel:2001yf,Martel:2003jj,Martel:2003th,Barack:2005nr}.
Here, we summarize the theory using a particular formulation that
makes the different expressions involved compact and self-explanatory.

We start from the perturbative splitting of the metric into the background,
the non-rotating Schwarzschild black hole metric $\met^{\sch}_{\alpha\beta}$, and the
{\em small} deviations $h_{\alpha\beta}$:
\begin{equation}
\met_{\alpha\beta} = \met^{\sch}_{\alpha\beta} + h_{\alpha\beta}\,. \label{schpert}
\end{equation}
Due to the spherical symmetry, the background manifold is the warped product
$M^2\times S^2$, where
$S^2$ denotes the 2-sphere and $M^2$ a two-dimensional Lorentzian
manifold.  The metric can then be written as the semidirect product of a
Lorentzian metric on $M^2$, $\met_{AB}$, and the unit curvature
metric on $S^2$, that we call $\gamma_{ab}$:
\begin{equation}
\met_{\alpha\beta} = \left(\begin{array}{cc}
\met_{AB} & 0 \\
0 & r^2\gamma_{ab} \end{array} \right) \,.\label{met22}
\end{equation}
Hereafter, $x^A$ denotes a coordinate system on $M^2$ and
$x^a$ a coordinate system on $S^2$; $r=r(x^A)$ is a function on $M^2$ that
coincides with the invariantly defined radial area coordinate.
In Schwarzschild coordinates we have
\begin{equation}
\met_{AB}dx^Adx^B = -f dt^2 +f^{-1}dr^2\,,~~
f = 1-2M/r\,.
\end{equation}
A vertical bar is used to denote the
covariant derivative on $M^2$ and a semicolon to denote the one on
$S^2$, thus we have $\met_{AB|C}=\gamma_{ab:c}=0\,.$   We can also introduce
the completely antisymmetric covariant unit tensors on $M^2$ and on $S^2$,
$\epsilon_{AB}$ and $\epsilon_{ab}$ respectively, in such a way they
satisfy:
$\epsilon_{AB|C}=\epsilon_{ab:c}=0\,,$
$\epsilon_{AC}\epsilon^{BC}= -\delta_A^B\,,$ and
$\epsilon_{ac}\epsilon^{bc}= -\delta_a^c\,.$
It is useful to introduce a vector field variable for the gradient of $r$:
\begin{equation}
w^{}_A\equiv r^{-1}\,r^{}_{|A}\,.
\end{equation}
Then, any covariant derivative on the spacetime can be written in
terms of the covariant derivatives on $M^2$ and $S^2$, plus some terms
due to the warp factor $r^2$, which can be written in terms of $w_A$.

Metric linear perturbations of a spherically-symmetric background can
be decomposed in scalar, vector and tensor spherical
harmonics~\cite{Gerlach:1979rw,Gerlach:1980tx}. The
scalar spherical harmonics $Y^{\ell m}$ are eigenfunctions of the
covariant Laplacian on the sphere:
\begin{equation}
\gamma^{ab}Y^{\ell m}_{:ab}=-l(l+1)Y^{\ell m}\,.
\end{equation}
A basis of vector spherical harmonics (defined for $l\ge 1$) is
\begin{equation}
Y^{\ell m}_a\equiv Y^{\ell m}_{:a}\,,~~~S^{\ell m}_a\equiv \epsilon_a{}^bY^{\ell m}_b\,,
\end{equation}
where the $Y_a^{\ell m}$'s have polar parity (they transform as $(-1)^l$,
like the scalar harmonics, under parity transformations, and are also called
even-parity type) and the
$S_a^{\ell m}$'s have axial parity (they transform as $(-1)^{l+1}$ under
parity transformations, and are also called odd-parity type). A basis of tensor
spherical harmonics (defined for $l\ge 2$) is
\begin{equation}
Y_{ab}^{\ell m}\equiv Y^{\ell m}\gamma_{ab}\,,~~~
Z^{\ell m}_{ab}\equiv Y^{\ell m}_{:ab}+\frac{l(l+1)}{2}Y^{\ell m}\gamma_{ab}\,,
\end{equation}
\begin{equation}
S^{\ell m}_{ab}\equiv S^{\ell m}_{(a:b)}\,,
\end{equation}
where the $Y_{ab}^{\ell m}\,,\,Z_{ab}^{\ell m}$ have polar parity and the
$S^{\ell m}_{ab}$ have axial parity.

We then split the metric perturbations $h_{\alpha\beta}$ into polar and axial
perturbations, $h_{\alpha\beta} = h^{\mbox{\small a}}_{\alpha\beta} +
h^{\mbox{\small p}}_{\alpha\beta}$, and these can be expanded in the basis of
tensor harmonics as
\begin{equation}
h^{\mbox{\small a}}_{\alpha\beta} =
\sum_{\ell,m} h^{\mbox{\small a},\ell m}_{\alpha\beta}\,,~~~
h^{\mbox{\small p}}_{\alpha\beta} =
\sum_{\ell,m} h^{\mbox{\small p},\ell m}_{\alpha\beta}\,,
\end{equation}
where
\begin{equation}
h^{\mbox{\small a},\ell m}_{\alpha\beta} = \left( \begin{array}{cc}
  0  &  q_A^{\ell m} \, S^{\ell m}_a \\
\\
 \ast  &  \  q_2^{\ell m} \, S_{ab}^{\ell m}
 \end{array}\right)\,, \label{maxial}
\end{equation}
\begin{equation}
h^{\mbox{\small p},\ell m}_{\alpha\beta} = \left( \begin{array}{cc}
  h_{AB}^{\ell m}\, Y^{\ell m} &  h_A^{\ell m} \, Y^{\ell m}_a \\
\\
  \ast  & \  r^2(K^{\ell m} \, Y^{\ell m}_{ab} +  G^{\ell m} \, Z_{ab}^{\ell m})
\end{array}\right)\,, \label{mpolar}
\end{equation}
and where we use asterisks to denote the symmetry of these tensors.  All the
perturbations, $h_{AB}^{\ell m}$ (scalar), $h_A^{\ell m}$ and $q_A^{\ell m}$ (vector),
and $K^{\ell m}\,$, $G^{\ell m}\,$, and $q_2^{\ell m}$ (tensor), depend only on
the coordinates of the 2-manifold $M^2$.

The components of the energy-momentum tensor of a point-like object are given by
\begin{equation}
T^{\alpha\beta} = m\int \frac{d\tau}{\sqrt{-g}}u^\alpha u^\beta
\delta^4[x-z(\tau)] \,, \label{tmunu}
\end{equation}
where $m$ is the mass, $\tau$ denotes proper time, $z(\tau)$ is the trajectory
of the object, $g$ denotes the metric determinant, and $\delta^4$ is the four-dimensional
Dirac density ($\int d^4x\sqrt{-g}\,\delta^4(x)=1$).  We choose to decompose in harmonics
the contravariant components $T^{\alpha\beta}$ [the decomposition for the
covariant ones follows immediately].  In this way, the polar components can be described
in terms of the following quantities
\begin{equation}
Q^{AB}_{\ell m} = 8\pi \int^{}_{S^2}\hspace{-3mm}d\Omega\,T^{AB}\bar{Y}^{\ell m}\,,
\end{equation}
\begin{equation}
Q^A_{\ell m} = \frac{16\pi r^2}{\ell(\ell+1)}\int^{}_{S^2}\hspace{-3mm}d\Omega\,T^{Aa}\bar{Y}^{\ell m}_a\,,
\end{equation}
\begin{equation}
Q^Y_{\ell m} = 8\pi r^2 \int^{}_{S^2}\hspace{-3mm}d\Omega\,T^{ab}\bar{Y}^{\ell m}_{ab}\,,
\end{equation}
\begin{equation}
Q^Z_{\ell m} = 32\pi r^4 \frac{(\ell-2)!}{(\ell+2)!}\int^{}_{S^2}\hspace{-3mm}d\Omega\,T^{ab}\bar{Z}^{\ell m}_{ab}\,,
\end{equation}
where the bar denotes complex conjugation.  The axial components can be described
in terms of the quantities
\begin{equation}
P^A_{\ell m} = \frac{16\pi r^2}{\ell(\ell+1)}\int^{}_{S^2}\hspace{-3mm}d\Omega\,T^{Aa}\bar{S}^{\ell m}_a\,,
\end{equation}
\begin{equation}
P^{}_{\ell m} = 16\pi r^4 \frac{(\ell-2)!}{(\ell+2)!}\int^{}_{S^2}\hspace{-3mm}d\Omega\,T^{ab}\bar{S}^{\ell m}_{ab}\,,
\end{equation}

To simplify the equations we choose to work in the Regge-Wheeler gauge:
\begin{equation}
h^{\ell m}_A = G^{\ell m}=0\,,~~
q^{\ell m}_2 = 0\,.
\end{equation}
but a fully covariant and gauge-invariant approach can be found in~\cite{Martel:2005ir}.

The perturbative equations can be decoupled by introducing certain
combinations of the metric perturbations, which are gauge-invariant.
For the axial modes, it is common to use the Regge-Wheeler master function
\begin{equation}
\Psi_{\ell m}^{\RW} =  -\frac{1}{r}w^Aq^{\ell m}_A \,,
\end{equation}
however, in this paper we will use the master function introduced
by Cunningham, Price and Moncrief~\cite{Cunningham:1978cp}, following
the definition used in~\cite{Jhingan:2002kb,Martel:2005ir}:
\begin{equation}
\Psi_{\ell m}^{\CPM} = \frac{2r}{(\ell+2)(\ell-1)}\epsilon^{AB}
\left(q^{\ell m}_{B|A}- \frac{2}{r}w^{}_Aq^{\ell m}_B \right)\,.
\end{equation}
One reason for using this function is to have the formulas for the
energy and angular momentum radiated to coincide with the ones for the
polar modes (see below).  In this respect, the formulation for axial
modes is on an equal footing with the one for polar modes.  For the polar modes,
we use the well-known Zerilli-Moncrief master function:
\begin{equation}
\Psi_{\ell m}^{\ZM} = \frac{2r}{\ell(\ell+1)}\left[K^{\ell m}+\frac{1}{\Lambda}
\left(h_{AB}^{\ell m}w^Aw^b-rw^AK^{\ell m}_{|A}\right)  \right]\,,
\end{equation}
where $\Lambda = (\ell+2)(\ell-1)/2 + 3M/r\,.$  The key point in introducing
these quantities is the fact that the Einstein perturbative equations can be decoupled for
them, and the remaining perturbative variables can be recovered from them.
As it is well know, the equations for $\Psi_{\ell m}^{\CPM}$ (or $\Psi_{\ell m}^{\RW}$)
and $\Psi_{\ell m}^{\ZM}$ are wave-type equations of the form:
\begin{equation}
\left[-\partial^2_t + \partial^2_{r^{}_{\!\ast}} - V^{\RW/\ZM}_\ell(r)\right]\Psi^{\CPM/\ZM}_{\ell m} =
f {\cal S}^{\CPM/\ZM}_{\ell m} \,,
\end{equation}
where $r^{}_{\!\ast}$ is the so-called {\em tortoise} coordinate ($r^{}_{\!\ast} =
r + 2M\ln(r/(2M)-1)$).  The potential for the axial modes is the Regge-Wheeler potential
\begin{equation}
V^{\RW}_\ell(r) = \frac{f}{r^2}\left( \ell(\ell+1)-\frac{6M}{r}\right)\,,
\end{equation}
and the one for polar modes is the Zerilli potential
\begin{equation}
V^{\ZM}_\ell(r) =\frac{f}{r^2\Lambda^2}\left[2\lambda^2_\ell\left(1+\lambda_\ell+
\frac{3M}{r}\right)+18\frac{M^2}{r^2}\left(\lambda_\ell+\frac{M}{r}\right)
\right]\,,
\end{equation}
where $\lambda_\ell = (\ell+2)(\ell-1)/2$.  The source term for axial modes is
given by
\begin{equation}
{\cal S}^{\CPM}_{\ell m} = \frac{2r}{\ell(\ell+1)}\epsilon^{AB}P^{\ell m}_{A|B}\,,
\end{equation}
and for polar modes by
\begin{widetext}
\begin{eqnarray}
{\cal S}^{\ZM}_{\ell m} & = & \frac{2}{\Lambda}w_A Q^A_{\ell m} - \frac{1}{r}Q^Z_{\ell m}
-\frac{r^2}{(1+\lambda^{}_\ell)\Lambda}\left\{w^C\met^{}_{AB}Q^{AB}_{\ell m|C} -
\frac{6M}{r^2\Lambda}w_Aw_BQ^{AB}_{\ell m} - \frac{f}{r}Q^Y_{\ell m} \right.\nonumber \\
 & - & \left. \frac{1}{r\Lambda}\left[\lambda^{}_\ell(\lambda^{}_\ell-1) +
3(2\lambda^{}_\ell-3)\frac{M}{r}+21\frac{M^2}{r^2}\right]\met^{}_{AB}Q^{AB}_{\ell m}
\right\}\,.
\end{eqnarray}
\end{widetext}
When we restrict ourselves to the case of a point particle, by introducing the
energy-momentum tensor (\ref{tmunu}) into the previous expressions, we find that
the source term for both polar and axial modes has the following (singular)
structure:
\begin{equation}
{\cal S}(t,r) = G(t,r)\delta[r-r^{}_p(t)] + F(t,r)\delta'[r-r^{}_p(t)]\,,
\label{sourcesform}
\end{equation}
where $\delta$ is the one-dimensional Dirac delta distribution and $\delta'$
its first derivative.  The function $r_p(t)$ describes the radial motion of the
particle in terms of the  coordinate time $t$.  In the polar case, the explicit
expressions for the functions $F(t,r)$ and $G(t,r)$ can be found, for instance,
in~\cite{Martel:2003th,Martel:2003jj}.  They are given by
\begin{eqnarray}
G^{\ZM}_{\ell m}(t,r) & = & a^{}_{\ell}(r)\bar{Y}^{\ell m}(t)
+ b^{}_{\ell}(r)\bar{Y}^{\ell m}_\varphi(t) \nonumber \\
& + & c^{}_{\ell}(r)\bar{Y}^{\ell m}_{\varphi\varphi}(t)
+ d^{}_{\ell}(r)\bar{Z}^{\ell m}_{\varphi\varphi}(t)\,,
\end{eqnarray}
\begin{eqnarray}
F^{\ZM}_{\ell m}(t,r) = A^{}_{\ell}(r)\bar{Y}^{\ell m}(t)\,,
\end{eqnarray}
where the $t$-dependence in the right-hand side has to be understood as:
$(t) = (\theta=\theta^{}_p=\pi/2,\varphi=\varphi^{}_p(t))$.  The different
functions of $r$ are given by:
\begin{widetext}
\begin{equation}
a^{}_{\ell}(r) = \frac{8\pi m}{1+\lambda^{}_\ell}\frac{f^2}{r\Lambda^2}\left\{
\frac{6M}{r}E^{}_p - \frac{\Lambda}{E^{}_p}\left[1+\lambda^{}_\ell
-\frac{3M}{r}+\frac{L^2_p}{r^2}\left(\lambda^{}_\ell+3-\frac{7M}{r}\right)
\right] \right\} \,,
\end{equation}
\end{widetext}
\begin{equation}
b^{}_{\ell}(r) = \frac{16\pi m}{1+\lambda^{}_\ell}\frac{L^{}_p}{E^{}_p}
\frac{f^2}{r^2\Lambda}u^r\,,
\end{equation}
\begin{equation}
c^{}_{\ell}(r) = \frac{8\pi m}{1+\lambda^{}_\ell}\frac{L^{2}_p}{E^{}_p}
\frac{f^3}{r^3\Lambda}\,,
\end{equation}
\begin{equation}
d^{}_{\ell}(r) = -32\pi m \frac{(\ell-2)!}{(\ell+2)!}\frac{L^{2}_p}{E^{}_p}
\frac{f^2}{r^3}\,,
\end{equation}
\begin{equation}
A^{}_{\ell}(r) = \frac{8\pi m}{1+\lambda^{}_\ell}\frac{f^3}{\Lambda}
\frac{1}{E^{}_p}\left(1+\frac{L^{2}_p}{r^2}\right)\,,
\end{equation}
where $E_p$ is the particle energy per unit mass, $L_p$ is the orbital
angular momentum, and $u^r$ is the radial component of the four-velocity,
which can be substituted by the expression given in equation~(\ref{eqmotion}).

We have computed the sources for the axial modes for the case in which
these perturbations are described by the Cunningham-Price-Moncrief master
function, since we are not aware of any reference in the literature 
containing them.  The result is the following:
\begin{equation}
G^{\CPM}_{\ell m}(t,r) = u^{}_{\ell}(r)\bar{S}^{\ell m}_\varphi(t)
+ v^{}_{\ell}(r)\bar{S}^{\ell m}_{\varphi\varphi} \,,
\end{equation}
\begin{equation}
F^{\CPM}_{\ell m}(t,r) = B^{}_{\ell}(r)\bar{S}^{\ell m}_\varphi(t)\,,
\end{equation}
where
\begin{widetext}
\begin{equation}
u^{}_{\ell}(r) = 32\pi m \frac{(\ell-2)!}{(\ell+2)!}\frac{f^2}{r^2}
\frac{L^{}_p}{E^{2}_p}\left[\left(1-\frac{5M}{r}\right)
\left(1+\frac{L^2_p}{r^2}\right) + f\frac{L^2_p}{r^2} - 2E^2_p\right]\,,
\end{equation}
\end{widetext}
\begin{equation}
v^{}_{\ell}(r) = 32\pi m \frac{(\ell-2)!}{(\ell+2)!}\frac{f^2}{r^3}
\frac{L^{2}_p}{E^{2}_p}u^r\,,
\end{equation}
\begin{equation}
B^{}_{\ell}(r) = 32\pi m \frac{(\ell-2)!}{(\ell+2)!}\frac{f^3}{r}
\frac{L^{}_p}{E^{2}_p}\left(1+\frac{L^2_p}{r^2}\right)\,,
\end{equation}
It is worth pointing out that for circular motion, there is only contribution 
from the vector harmonics.

The singular structure of the source terms in the master equations, for both
polar and axial modes, shown in equation~(\ref{sourcesform}) implies the existence
of discontinuities in the master function at the particle's location. Outside
the location of the particle the solution is smooth (assuming it was initially).
One can then divide the one-dimensional spatial domain into two different and
disjoint regions (which will change in time as the particle moves): The region to the left
of the particle ($r<r^{}_p(t)$), and the region to the right of the particle
($r>r^{}_p(t)$).   The master functions, on that regions, satisfy an homogeneous
wave-like equation [Equation~(\ref{equation}) without sources].  Then, we can think
of the solution of the full equation as being composed of the solutions of the
homogeneous equations to the left and to the right of the particle, and relations
between them at the particle location.  Obviously these relations consist of
imposing the discontinuities that the singular source terms dictate.
In other words, we can write the solution of the full equation as
\begin{equation}
\Psi(t,r) = \Psi^-_{}(t,r)\theta(r^{}_p(t)-r) + \Psi^+_{}(t,r)\theta(r-r^{}_p(t))\,,
\label{distribution}
\end{equation}
where $\theta(r)$ is the Heaviside step function, and $\Psi^+_{}(t,r)$ and
$\Psi^-_{}(t,r)$ are solutions of the homogeneous equation to the right and
to the left of the particle respectively.  Due to the existence
of the particle we will have that
\begin{equation}
\Psi^-_{}(t,r^{}_p) \neq \Psi^+_{}(t,r^{}_p)\,,~~~
(\partial^{}_r\Psi^-_{})(t,r^{}_p) \neq (\partial^{}_r\Psi^+_{})(t,r^{}_p)\,.
\end{equation}
By introducing~(\ref{distribution}) into the full equation we can derive the
equation that govern the jumps in the master functions.  The result is:
\begin{eqnarray}
\left(f^2(t)-\dot{r}^{2}_p(t)\right)\left[\partial^{}_r\Psi\right] - 2\,\dot{r}^{}_p(t)
\partial^{}_t\,\left[\Psi\right] \nonumber \\ 
- \left(\ddot{r}^{}_p(t)+f(t)f'(t)\right)\left[\Psi\right] = 
{\cal G}(t,r^{}_p(t)) \,, \label{discon1}
\end{eqnarray}
\begin{equation}
\left(f^2(t)+\dot{r}^{2}_p(t)\right)\left[\Psi\right] = {\cal F}(t,r^{}_p(t)) \,,\label{discon2}
\end{equation}
where $f(t)\equiv f(r^{}_p(t))$, and $\left[\Psi\right]$ and $\left[\partial^{}_r\Psi\right]$ 
are the master function discontinuities at the particle location
\begin{equation}
\left[\Psi\right] = \lim^{}_{r\rightarrow r^{}_p(t)}\Psi^+_{}(t,r) -
\lim^{}_{r\rightarrow r^{}_p(t)}\Psi^-_{}(t,r)\,,
\end{equation}
\begin{equation}
\left[\partial^{}_r\Psi\right] = \lim^{}_{r\rightarrow r^{}_p(t)}\partial^{}_r\Psi^+_{}(t,r) -
\lim^{}_{r\rightarrow r^{}_p(t)}\partial^{}_r\Psi^-_{}(t,r)\,,
\end{equation}
where ${\cal F}$ and ${\cal G}$ are functions of $t$ and the particle location
$r^{}_p(t)$ that one obtains after applying the following
properties of the Dirac delta distribution
\begin{equation}
A(t,r)\delta(r-r^{}_p(t)) = A(t,r^{}_p(t))\delta(r-r^{}_p(t)) \,,
\end{equation}
\begin{eqnarray}
A(t,r)\delta'(r-r^{}_p(t)) & = & -(\partial^{}_rA)(t,r^{}_p(t))\delta(r-r^{}_p(t))
\nonumber \\
& + & A(t,r^{}_p(t))\delta'(r-r^{}_p(t))\,,
\end{eqnarray}
to the original source terms~(\ref{sourcesform}).  The equations for the
discontinuities [Equations~(\ref{discon1}) and~(\ref{discon2})] contain the particle
radial position $r^{}_p(t)$,
and its first and second time derivatives $\left(\dot{r}^{}_p(t),\ddot{r}^{}_p(t)\right)$.
The first two, $r^{}_p(t)$ and $\dot{r}^{}_p(t)$, are obtained via
numerical integration of the geodesics ODEs shown in Appendix~\ref{geodesic};
and the third one, $\ddot{r}^{}_p(t)$, can be found directly from the geodesic equations:
\begin{eqnarray}
\ddot{r}^{}_p(t) & = & \frac{f^2(t)}{r^2_p(t)}\frac{L^2_p}{E^2_p}\left(
\frac{f(t)}{r^{}_p(t)}-\frac{3f'(t)}{2}\right) \nonumber \\
& + &  f(t)f'(t)\left(1-\frac{3f^2(t)}{2E^2_p}\right)  \,.
\end{eqnarray}

We finish this section by giving the expressions of the averaged energy at angular momentum 
luminosities at infinity (as obtained from the Isaacson's averaged energy-momentum tensor 
for gravitational waves~\cite{Isaacson:1968ra,Isaacson:1968gw}), which also hold at the horizon, 
in terms of the axial and polar master functions:
\begin{equation}
\dot{E} = \frac{1}{64\pi}\sum^{}_{\ell\geq 2, m}\frac{(\ell+2)!}{(\ell-2)!}\left(
|\dot{\Psi}^{\CPM}_{\ell m}|^2 + |\dot{\Psi}^{\ZM}_{\ell m}|^2\right)\,,
\end{equation}
\begin{equation}
\dot{L} = \frac{1}{64\pi}\sum^{}_{\ell\geq 2, m}i m\frac{(\ell+2)!}{(\ell-2)!}\left(
\bar{\Psi}^{\CPM}_{\ell m}\dot{\Psi}^{\CPM}_{\ell m} +
\bar{\Psi}^{\ZM}_{\ell m}\dot{\Psi}^{\ZM}_{\ell m}\right)\,.
\end{equation}
Finally, the metric {\em waveforms} are given by
\begin{equation}
h^{}_{+} - i h^{}_{\times} = \frac{1}{2r}\sum^{}_{\ell\geq 2, m}
\sqrt{\frac{(\ell+2)!}{(\ell-2)!}}\left\{\Psi^{\ZM}_{\ell m}+ i \Psi^{\CPM}_{\ell m}
\right\}{}^{}_{-2}Y^{\ell m} \,,
\end{equation}
where ${}^{}_{-2}Y^{\ell m}$ are the spherical harmonics of spin weight $-2$
(see, e.g.~\cite{Goldberg:1967sp}).

\section{The numerical framework}\label{femformulation}
In this section we introduce the basics on the numerical framework that we use
to solve numerically the perturbative equations described above.
This task involves a number of choices that determine the particular
features of our numerical method.  To be specific, we want to solve our
equations in the {\em time domain}, that is, we want to develop an
algorithm that evolves the initial data from an initial state to a final
time where we are interested in knowing the solution.  This implies a
discretization of our equations both in space and time.  We choose
to discretize in space by using a Galerkin-type FE procedure, and in time by using
Finite Differences techniques.  In what follows we describe the details of these
ingredients of our numerical calculations.

\subsection{The Mathematical Formulation of the Problem}\label{maths}
The model PDE problem that we are interesting in solving has the following form:
\begin{equation}
{\cal L}[\Psi] \equiv \left(-\partial^2_t + \partial^2_x -V(x)\right)\Psi(t,x) -
{\cal S}(t,x) = 0\,, \label{equation}
\end{equation}
\begin{equation}
\Psi(t_o,x) = \psi_o(x)\,,~~~~
(\partial_t\Psi)(t_o,x) = \dot\psi_o(x)\,, \label{initialdata}
\end{equation}
\begin{equation}
[(\partial_t - \partial_x)\Psi](t,x^{}_H) = 0\,,~~~~
[(\partial_t + \partial_x)\Psi](t,x^{}_I) = 0\,, \label{boundaries}
\end{equation}
where
\begin{equation}
t\in[t_o,t_f]\,,~~\mbox{and}~~ x\in \Omega = [x^{}_H,x^{}_I]\,. \label{domains}
\end{equation}
Expression~(\ref{equation}) presents the structure of the equations we need to solve,
namely a wave equation in a potential $V$ and with a source ${\cal S}$.  Here,
$x$ corresponds to the tortoise coordinate $r^{}_{\!\ast}$ in the master equations,
and for the sake of simplicity of the notation we will use it in most of the
rest of the paper.   The form of ${\cal S}(t,x)$ is assumed to be known,
as it is in our case.  Equation (\ref{initialdata}) represents the initial
conditions for the time evolution, we need the initial value of
$\Psi$ and its time derivative to solve a second-order hyperbolic problem.
The boundary conditions, given in (\ref{boundaries}), are simple one-dimensional
outgoing conditions (also known as Sommerfeld boundary conditions) at both ends of the spatial
interval where the equation is considered, specified in~(\ref{domains}).
This boundary condition is an exact outgoing boundary condition only at infinity,
provided the potential and the source have a fall-off of the type of the potentials 
and sources that we are considering in our physical problem, otherwise it is just 
an approximate outgoing
boundary condition whose accuracy depends on how far $x^{}_I$ is located.
A similar argument also holds for the boundary condition at $x^{}_H$, the
accuracy of the boundary condition that we use depends on how far towards
$-\infty$ we locate $x^{}_H$.   There is a well-known difference between
$x\rightarrow\infty$ and $x\rightarrow -\infty$ which is due to the asymptotic
behaviour of the potentials.  When we approach $\rightarrow\infty$ the potentials
behave like $const.\times x^{-2}$, whereas when we approach $\rightarrow -\infty$
the potentials behave like $const.\times\exp\{-x/(2M)\}$, therefore our
approximate boundary conditions should work much better at $x^{}_H$ than at
$x^{}_I$ since our equations resemble the wave equation better around $x^{}_H$.
One could also get better boundary conditions by using the methods suggested
in references~\cite{Bayliss:1980bt,Bayliss:1982bg}, where higher-order
derivative boundary conditions are proposed, or by using the methods proposed
in~\cite{Alpert:2000bt,Alpert:2002bt,Lau:2004jn,Lau:2004as,Lau:2005ti}
where {\em exact} radiative boundary conditions are studied.

\subsection{The Finite Element Discretization}\label{femdiscrete}
In this section we describe, in a simplified way, the main ingredients of the FEM
that are relevant for our calculations.  Detailed exposions can
be found in~\cite{StrFix:73,Hughes:1987tj,Babuska:2001bs,Zienkiewicz:77oc}.

The way one discretizes in space in a FE framework can be summarized
in the following three steps: (i) {\em Domain discretization.} The division of the
spatial domain $\Omega$ into a collection of disjoint subdomains
$\{\Omega^{}_k\}^{}_{i=1\ldots N}\,,$ the {\em elements}.
(ii) {\em The FE functional space}.  At every element,
$\Omega_k$, we introduce a finite-dimensional functional space, ${\cal F}_k$,
that we use to expand our fields locally at $\Omega_k\,.$  Typically, these
functional spaces are made out of polynomials.
(iii) {\em Weak formulation of the equation}.  This consists in converting
the differential form of the equations into an integral form that involves
the boundary conditions of the von Neumann type.
(iv) {\em Equation discretization.} In a Garlekin-type of FE formulation, the
discretized equations are obtain throught the imposition of the vanishing of all
the {\em residuals}, ${\cal E}_A \equiv \int_\Omega dx\, n^{}_i\, {\cal L}[\Psi]\,,$
the components of our equation with respect a basis of nodal functions
$\{n^{}_i(x)\}$ built out of the spaces ${\cal F}_k$ (see below).

\subsubsection{Domain Discretization and the FE functional spaces}
Points (i) and (ii) have to be treated jointly, because the structure of
the elements and the structure of the FE functional spaces are not
completely independent.

\begin{figure}[htb]
\begin{center}
\includegraphics[width=85mm]{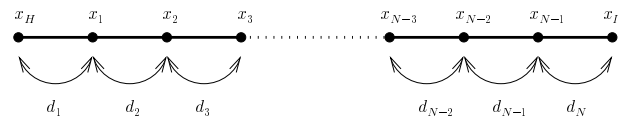}
\caption{One-dimensional Mesh. \label{mesh_ini}}
\end{center}
\end{figure}

Since we are dealing with a one-dimensional problem,
the first step, the subdivision of the domain, is quite simple, we just
divide the interval $[x^{}_H,x^{}_I]$ into subintervals (see Figure~\ref{mesh_ini}):
$\Omega^{}_1=[x^{}_H,x^{}_1)\,,$ $\Omega^{}_2=[x^{}_1,x^{}_2)\,,$ \ldots\,,
$\Omega^{}_N=[x^{}_{N-1},x^{}_I]\,.$  This equivalent to locate $N+1$ nodes
in an ordered way: $x^{}_0\equiv x^{}_H < x^{}_1 < \cdots < x^{}_{N-1}
< x^{}_N\equiv x^{}_I\,.$  We denote the size (length) of the element
$\Omega^{}_k$ by $d^{}_k = x^{}_{k+1} - x^{}_k$.

\begin{figure}[htb]
\begin{center}
\includegraphics[width=70mm]{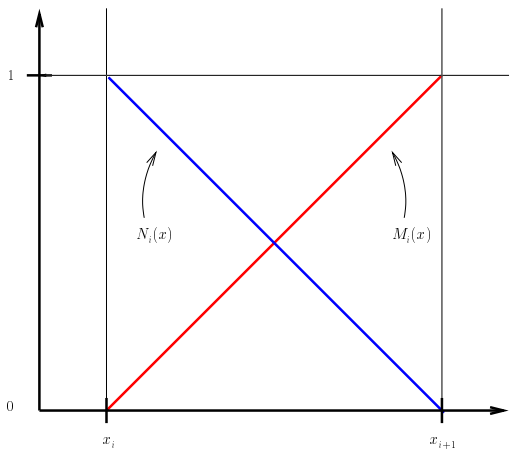}
\caption{Linear interpolation functions $M^{}_i(x)$ and  $N^{}_i(x)\,.$
\label{functional}}
\end{center}
\end{figure}

\begin{figure}[htb]
\begin{center}
\includegraphics[width=85mm]{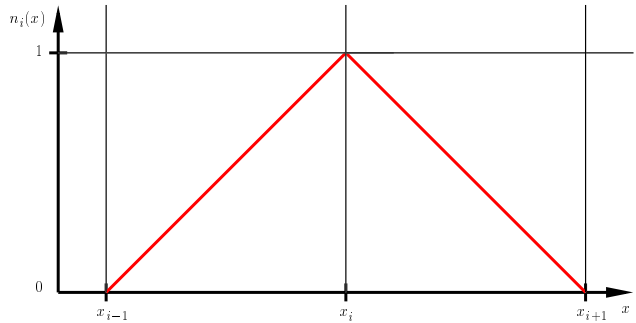}
\caption{Nodal functions $n^{}_i(x)\,.$ \label{nodal}}
\end{center}
\end{figure}

The second step is very important in the sense that the accuracy and
convergence properties of the FE scheme depend on the choice of
the FE functional space.  In this paper, we restrict ourselves to linear elements.
For sufficiently regular meshes, linear elements lead to second-order
convergence in the $L^2$ norm.  On the other hand, the functional spaces ${\cal F}_k$
for linear elements are two-dimensional and can be span by the
following two functions (see Figure~\ref{functional}):
\begin{equation}
M^{}_i(x) = \left\{ \begin{array}{ll}
\frac{x-x_i}{d^{}_i} & \mbox{{}~if $x\in (x_i,x_{i+1})$,} \\
0 & \mbox{{}~otherwise,}
\end{array}  \right.
\end{equation}
\begin{equation}
N^{}_i(x) = \left\{ \begin{array}{ll}
\frac{x_{i+1}-x}{d^{}_i} & \mbox{{}~if $x\in (x_i,x_{i+1})$,} \\
0 & \mbox{{}~otherwise,}
\end{array}  \right.
\end{equation}
which are usually called {\em linear interpolation} functions. However,
to build a FE approximation of the solution of our PDEs it is more convenient
to use the {\em nodal} functions, $n^{}_i(x)$:
\begin{equation}
n^{}_H(x) = N^{}_H(x)\,,~~~
n^{}_I(x) = M^{}_{N-1}(x)\,.
\end{equation}
and for $i=1,...,N-1$:
\begin{equation}
n^{}_i(x) = \left\{ \begin{array}{ll}
M^{}_{i-1}(x) & \mbox{{}~if $x\in (x_{i-1},x_i)$,} \\
N^{}_i(x) & \mbox{{}~if $x\in (x_i,x_{i+1})$,} \\
0 & \mbox{{}~otherwise.}
\end{array}  \right.
\end{equation}
Their are called nodal functions because of the following
property (see Figure~\ref{nodal}):
\begin{equation}
n^{}_i(x_j) = \delta_{ij}\,,
\end{equation}
that is, they vanish at all nodes excepting at the one they are
associated with, where they take the unity value.

\subsubsection{The weak form of the equations}
The next step is to formulate our problem in what is called the {\it weak} form
of the equation, which is an integral form.  To that end, let us consider an arbitrary
{\em test} function $\phi$ on
$[t_o,t_f]\times [x^{}_H,x^{}_I]$ and multiply equation~(\ref{equation})
by it.  Then, we integrate over $[x^{}_H,x^{}_I]$ and apply integration
by parts to the term with second spatial derivatives.  This produces
a boundary term with first spatial derivatives that can be converted into
first time derivatives by using our boundary conditions~(\ref{boundaries}).
The result is:
\begin{widetext}
\begin{equation}
{\cal E}[\phi,\Psi] = -\int^{x^{}_I}_{x^{}_H}dx\left[\phi\partial^2_t\Psi
+ (\partial_x\phi)(\partial_x\Psi) +V(x)\phi\Psi \right]
+ \left.\phi\,\partial_t\Psi\right|_{x^{}_I}
+ \left.\phi\,\partial_t\Psi\right|_{x^{}_H} -
\int^{x^{}_I}_{x^{}_H}dx\, \phi\, {\cal S} = 0 \,. \label{weak}
\end{equation}
\end{widetext}
This is the {\em weak} form of our equation, which has the remarkable
property of including the boundary conditions of the problem.  In the
case we were dealing with Dirichlet boundary conditions, usually called
{\em essential} boundary conditions in the FEM language, they would have not been
incorporated in the {\em weak} formulation of the equation.   Instead, they are used
to eliminate unknowns by providing values for them.

\subsubsection{The FE discretization of the equation}
To find the FE discretization of our equations we need to introduce
the FE approximation to the solution of our problem.  This approximation consists
in an expansion in the nodal functions
\begin{equation}
\Psi_h(t,x) = \sum_{i=0}^{N} \psi^{}_i(t) n^{}_i(x) \,, \label{feaprox}
\end{equation}
where the time-dependent functions $\psi^{}_i(t)$ are going to
be the unknowns of our problem.  Then, the FE discretization of the
equations of our problem consists in imposing the vanishing of the
{\em residuals}:
\begin{equation}
{\cal E}_i \equiv {\cal E}[n^{}_i,\Psi_h] = 0\,,
~~~i=0\,,\,...\,,\,N\,.  \label{residuals}
\end{equation}
This leads to one equation for node\footnote{In the case of problems involving
{\em essential} boundary conditions we would get as many equations as
unknows remain after imposing these boundary conditions.}, or equivalently, we have $N+1$
equations for our $N+1$ unknowns $\{\psi_i\}\,.$  By introducing the
FE approximation (\ref{feaprox}) into the {\em weak} form of our equation
(\ref{weak}) and imposing (\ref{residuals}), we get a system of coupled ODEs for
the unknowns.  We can write it in a matrix form as follows:
\begin{equation}
\mathbb{M}\,\mb{\cdot}\mb{\ddot{\Psi}} + \mathbb{G}\,\mb{\cdot}\mb{\dot{\Psi}} +
\mathbb{K}\,\mb{\cdot}\mb{\Psi} = \mb{F}\,, \label{femodes}
\end{equation}
where $\mathbb{M}$, $\mathbb{G}$, and $\mathbb{K}$ are $(N+1)\times(N+1)$
matrices, and $\mb{\Psi}$ and $\mb{F}$ are vectors with $N+1$ components, and
$\mb{\ddot{\Psi}}$ and $\mb{\dot{\Psi}}$ are the time derivatives of $\mb{\Psi}$.
The names and meaning of these objects can be thought to be inspired in the
analogy of the system of equations (\ref{femodes}) with the system of equations
for a coupled system of oscillators.  The matrix $\mathbb{M}$ is usually called
the {\em mass} matrix and its components are:
\begin{equation}
\mathbb{M}_{ij} = \int^{x^{}_I}_{x^{}_H}dx\, n^{}_i(x) n^{}_j(x) \,,
\end{equation}
and hence it is a symmetric and positive definite matrix.  The matrix $\mathbb{G}$
is usually called the {\em damping matrix} and has components
\begin{equation}
\mathbb{G}_{ij} = n^{}_i(x^{}_H)n^{}_j(x^{}_H) + n^{}_i(x^{}_I)n^{}_j(x^{}_I) \,,
\end{equation}
which is symmetric and only has contributions from the boundaries, which means
that our system only dissipates (or absorbs, if they sign of these two terms would
have been negative, corresponding to ingoing boundary conditions) 
{\em energy}\footnote{We can define an energy for our system by:\\
\[
\mbox{E}[\mb{\Psi},\mb{\dot{\Psi}}] = \frac{1}{2}\left( \mb{\dot{\Psi}}^{\mbox{\tiny T}}
\mb{\cdot}\mathbb{M}\mb{\cdot}\mb{\dot{\Psi}} + \mb{\Psi}^{\mbox{\tiny T}}\mb{\cdot}
\mathbb{K}\,\mb{\cdot}\mb{\Psi}\right)\,.\]
This energy would be preserved by the evolution when $\mathbb{G} = \mb{F} = 0\,.$}
through the boundaries and this reflects the fact that we are using outgoing boundary
conditions.
The matrix $\mathbb{K}$ is usually called the {\em stiffness} matrix.  Its components
are given by
\begin{equation}
\mathbb{K}_{ij} = \int^{x^{}_I}_{x^{}_H}dx\,\left[ n'_i(x) n'_j(x) +
 V(x) n^{}_i(x) n^{}_j(x)\right] \,, \label{stiffness}
\end{equation}
therefore, it is also symmetric.  Finally, $\mb{\Psi} = (\psi_i(t))$ and
$\mb{F}$ is usually called the {\em force} vector and its components are given by
\begin{equation}
\mb{F}_i(t) = - \int^{x^{}_I}_{x^{}_H}dx\, {\cal S}(t,x)n^{}_i(x)\,.
\end{equation}

In our particular case, we can compute most of the components of the matrices
and vectors that determine our system of ODEs (\ref{femodes}) analytically.
The expressions for the components of {\em mass} matrix are:
\begin{equation}
\mathbb{M}_{ij} = \textstyle{\frac{1}{6}}\left[d^{}_{i-1}\delta^{}_{i-1\,j}
+ 2\left(d^{}_{i-1} + d^{}_{i}\right)\delta^{}_{ij} +
d^{}_i \delta^{}_{i+1\,j}  \right]\,,
\end{equation}
for $i,j,=1,\ldots,N-1$, and the components related to the boundaries are:
$\mathbb{M}_{HH} = d^{}_H/3\,,$
$\mathbb{M}_{H\,1} = d^{}_H/6\,,$
$\mathbb{M}_{N-1\,I} = d^{}_{N-1}/6\,,$ and
$\mathbb{M}_{II} = d^{}_{N-1}/3\,.$
The components of the {\em damping} matrix are simply given by
\begin{equation}
\mathbb{G}_{ij} = \delta^{}_{i\,H}\delta^{}_{j\,H}
+ \delta^{}_{i\,I}\delta^{}_{j\,I}  \,,
\end{equation}
The first term of the components of the {\em stiffness} matrix is given by
\begin{equation}
\mathbb{K}_{ij} = -\frac{1}{d^{}_{i-1}}\delta^{}_{i-1\,j}
+ \left(\frac{1}{d^{}_{i-1}} + \frac{1}{d^{}_{i}}\right)
\delta^{}_{ij} - \frac{1}{d^{}_{i-1}}\delta^{}_{i+1\,j} \,,
\end{equation}
for $i,j,=1,\ldots,N-1$, and the components related to the boundaries are:
$\mathbb{K}_{HH} = 1/d^{}_H\,,$
$\mathbb{K}_{H\,1} = -1/d^{}_H\,,$
$\mathbb{K}_{N-1\,I} = -1/d^{}_{N-1}\,,$ and
$\mathbb{K}_{II} = -1/d^{}_{N-1}\,.$
The second term in the components of the {\em stiffness} matrix involves the 
potential, and therefore it has to be computed numerically.  To that end, we 
use Gauss-Legendre quadratures (see Appendix~\ref{GaussLegendre}).

We can compute the components of the {\em force} vector $\mb{F}$ by using the
form of the source term ${\cal S}\,,$ which is given in equation~(\ref{sourcesform}).
Using the properties of the Dirac delta distribution, we find that the 
structure of ${\cal S}$ implies the following structure for the components 
of the {\em force} vector:
\begin{eqnarray}
\mb{F}^{}_i(t) & = & \left\{ \nu(r^{}_p(t))\left[(\partial_rF)(t,r^{}_p(t))
- G(t,r^{}_p(t)) \right] \right. \nonumber \\
& + & \left. \nu'(r^{}_p(t))F(t,r^{}_p(t))\right\}n^{}_i(x^{}_p(t)) \nonumber \\
& + & \nu^2(r^{}_p(t))F(t,r^{}_p(t))n'^{}_i(x^{}_p(t)) \,,
\end{eqnarray}
where
\begin{equation}
\nu(r) = \frac{dx(r)}{dr} = \frac{1}{f}\,,~~
\nu'(r) = \frac{d\nu(r)}{dr} = -\frac{2M}{r^2f^2}\,,
\end{equation}
and $x_p(t)$ is just the radial motion in terms of the {\em tortoise}
coordinate.  This completes the FE discretization of our
problem.

\subsection{Evolution Algorithms}
The next step is to solve the system of ODEs given in equation~(\ref{femodes}),
which is coupled to the ODEs corresponding to the motion of the point-like object
[equations (\ref{ode1},\ref{ode2}) in Appendix~\ref{geodesic}].  The numerical
algorithms we use derive from
the {\em average acceleration} method, which is based on the assumption that over
a small time interval any nodal acceleration can be considered to be a linear
function of time.   Then, for a time interval $(t_o,t_o+\Delta t)$, we write
\begin{equation}
\mb{\ddot{\Psi}}(t) = \mb{\ddot{\Psi}}(t_o)\left(1-\frac{t}{\Delta t}\right)
+ \mb{\ddot{\Psi}}(t_o+\Delta t) \frac{t}{\Delta t}\,.
\end{equation}
Integrating in time this equation twice and evaluating at
$t= t_1 = t_o+\Delta t$ we get
\begin{equation}
\mb{\dot{\Psi}}(t_1) = \mb{\dot{\Psi}}(t_o)+\frac{1}{2}\left[
\mb{\ddot{\Psi}}(t_o) +\mb{\ddot{\Psi}}(t_1)\right]\Delta t\,,
\label{speed0}
\end{equation}
\begin{equation}
\mb{\Psi}(t_1) = \mb{\Psi}(t_o)+\mb{\dot{\Psi}}(t_o)\Delta t +
\frac{1}{6}\left[2\mb{\ddot{\Psi}}(t_o) +\mb{\ddot{\Psi}}(t_1)\right]
(\Delta t)^2\,. \label{acceleration0}
\end{equation}
This algorithm that one derives from these expressions is conditionally stable.  
Newmark~\cite{Newmark:1959nn} introduced a generalization of the equations 
(\ref{speed0}) and (\ref{acceleration0}) in the following way
\begin{equation}
\mb{\dot{\Psi}}(t_1) = \mb{\dot{\Psi}}(t_o)+
\left[(1-\gamma)\mb{\ddot{\Psi}}(t_o) +
\gamma\mb{\ddot{\Psi}}(t_1)\right]\Delta t\,,\label{speed1}
\end{equation}
\begin{eqnarray}
\mb{\Psi}(t_1) & = & \mb{\Psi}(t_o)+\mb{\dot{\Psi}}(t_o)\Delta t
\nonumber \\
& + & \frac{1}{2}\left[\left(1-2\beta\right)\mb{\ddot{\Psi}}(t_o) +
2\beta\mb{\ddot{\Psi}}(t_1)\right](\Delta t)^2\,,\label{acceleration1}
\end{eqnarray}
where $(\gamma,\beta)$ are parameters that have to be chosen for
accuracy and stability.  The Newmark method is unconditionally
stable for the following range of the parameters $(\gamma,\beta)$:
\begin{equation}
\gamma \geq \frac{1}{2}\,,~~~
\beta \geq \frac{1}{4}\left(\frac{1}{2}+\gamma \right)^2\,.
\end{equation}
For $(\gamma,\beta)=(1/2,1/6)$ we recover the average acceleration
method; for $(\gamma,\beta)=(1/2,0)$ we obtain the central differences
method (although it is not strictly explicit), which is conditionally
stable; and for $(\beta,\gamma)=(1/4,1/2)$ we get the {\em trapezoidal}
rule, which is unconditionally stable and second-order accurate.
In the Newmark scheme, equation (\ref{femodes}) is left untouched.

However, numerical damping to prevent the amplification of high-frequency
modes cannot be introduced in the Newmark algorithm without degrading
the order of accuracy to first-order.
There are a  number of numerical schemes that generalize
the Newmark scheme in order to include maximal dissipation of high
frequency modes and minimal of low frequency modes and at the same
time maintaining second-order accuracy.   In particular: the Hilber-$\alpha$
method~\cite{Hilber:1977hi}, the Bossak-$\alpha$ method~\cite{Wood:1981bo},
and the Generalized-$\alpha$ method~\cite{Chung:1993ge}.  We present
here the last one, which includes, for certain values of the parameters,
the other methods.  The Generalized-$\alpha$ method can be seen as a
generalization of Newmark's algorithm in the sense that equations
(\ref{speed1},\ref{acceleration1}) are also assumed by this evolution
scheme.  The generalization takes place when we discrete the set of ODEs
given in equation~(\ref{femodes}).  Let
$(\mb{\Psi}^{}_n,\mb{\dot{\Psi}}^{}_n,\mb{\ddot{\Psi}}^{}_n)$ be the values
of our unknowns and their time derivatives at a time $t=t^{}_n$.  Then, the
discretization of (\ref{femodes}) used in the Generalized-$\alpha$
method is given by
\begin{equation}
\mathbb{M}\mb{\cdot}\mb{\ddot{\Psi}}^{}_{n+1-\alpha^{}_m}
+ \mathbb{G}\mb{\cdot}\mb{\dot{\Psi}}^{}_{n+1-\alpha^{}_f}
+ \mathbb{K}\mb{\cdot}\mb{\Psi}^{}_{n+1-\alpha^{}_f} =
\mb{F}^{}_{n+1-\alpha^{}_f}\,, \label{generalized}
\end{equation}
where
\begin{equation}
\mb{\ddot{\Psi}}^{}_{n+1-\alpha^{}_m} = (1-\alpha^{}_m)
\mb{\ddot{\Psi}}^{}_{n+1} + \alpha^{}_m\mb{\ddot{\Psi}}^{}_{n}\,,
\end{equation}
\begin{equation}
\mb{\dot{\Psi}}^{}_{n+1-\alpha^{}_f} = (1-\alpha^{}_f)
\mb{\dot{\Psi}}^{}_{n+1} + \alpha^{}_f\mb{\dot{\Psi}}^{}_{n}\,,
\end{equation}
\begin{equation}
\mb{\Psi}^{}_{n+1-\alpha^{}_f} = (1-\alpha^{}_f)
\mb{\Psi}^{}_{n+1} + \alpha^{}_f\mb{\Psi}^{}_{n}\,,
\end{equation}
\begin{equation}
\mb{F}^{}_{n+1-\alpha^{}_f} = (1-\alpha^{}_f)
\mb{F}^{}_{n+1} + \alpha^{}_f\mb{F}^{}_{n} \,,
\end{equation}
where $\alpha^{}_f$ and $\alpha^{}_m$ are constants.  The Newmark
method corresponds to $(\alpha^{}_f,\alpha^{}_m)=(0,0)$, the 
Hilber$-\alpha$ method to $\alpha^{}_m = 0$, and the Bossak$-\alpha$
method to $\alpha^{}_f = 0\,.$
Introducing equations (\ref{speed1}) and (\ref{acceleration1}) into equation
(\ref{generalized}) and rearranging the different terms, we arrive at the following
equation for $\mb{\ddot{\Psi}}^{}_{n+1}$:
\begin{widetext}
\begin{align}
& \left[ (1-\alpha^{}_m)\mathbb{M} + (1-\alpha^{}_f)\gamma\Delta t\,
\mathbb{G} + (1-\alpha^{}_f)\beta(\Delta t)^2 \mathbb{K} \right]\mb{\cdot}
\mb{\ddot{\Psi}}^{}_{n+1} = (1-\alpha^{}_f)
\mb{F}^{}_{n+1} + \alpha^{}_f\mb{F}^{}_{n} -
\alpha^{}_m\mathbb{M}\mb{\cdot}\mb{\ddot{\Psi}}^{}_{n}
\nonumber \\
&\hspace{2cm} - \mathbb{G}\mb{\cdot}\left[\mb{\dot{\Psi}}^{}_{n}
+(1-\alpha^{}_f)(1-\gamma)\Delta t\,\mb{\ddot{\Psi}}^{}_{n}\right] -
\mathbb{K}\mb{\cdot}\left\{ \mb{\Psi}^{}_{n} + (1-\alpha^{}_f)\Delta t\,
\left[\mb{\dot{\Psi}}^{}_{n}+ (\frac{1}{2}-\beta)\Delta t\,\mb{\ddot{\Psi}}^{}_{n}
\right] \right\}\,. \label{mastereq}
\end{align}
\end{widetext}
Then, the algorithm that we are going to use to solve these equations
for our unknowns goes as follows: (i) We solve (\ref{mastereq}) for 
$\mb{\ddot{\Psi}}^{}_{n+1}$,
(ii) We compute $\mb{\dot{\Psi}}^{}_{n+1}$ from (\ref{speed1}) and,
(iii) We compute $\mb{\Psi}^{}_{n+1}$ from (\ref{acceleration1}).
Excepting in very special cases, the method that comes out of this
algorithm is implicit.   In general implicit schemes are computationally
expensive, but since we are using one-dimensional
linear elements the matrices that we are dealing with are
symmetric tridiagonal and therefore, one can use fast routines to
invert them (see, e.g.~\cite{Press:1992nr}).

The convergence and stability properties of these algorithms and
their high-frequency damping capabilities can be analyzed by casting the 
time discretization of our ODEs in the form: $\mb{U}^{}_{n+1} = \mathbb{A}
\mb{\cdot}\mb{U}_n + \mb{R}_n\,,$ and then to analyze the truncation
error when $\mb{U}_n$ is substituted by the exact expression and
the spectral properties of the so-called {\em amplification} matrix
$\mathbb{A}$ (see, e.g.~\cite{Hughes:1987tj}).   A quantity that plays an
important role is the spectral radius
\begin{equation}
\rho_\infty = \lim^{}_{\Delta t/T\rightarrow\infty} \rho(\mathbb{A})\,,
\end{equation}
where $\rho(\mathbb{A})$ is the spectral radius of the matrix $\mathbb{A}$,
$\Delta t$ is the time step, and $T$ is the {\em vibration} period of a
generic mode of the system.  For $\rho_\infty=1$ there is no damping, and
the lower $\rho_\infty$ is, the bigger the damping gets.
In table~\ref{parameters} we show the values of the time integration
parameters $(\alpha^{}_m,\alpha^{}_f,\beta,\gamma)$ for optimal damping
properties in terms of the spectral radius $\rho^{}_\infty$
(see~\cite{Hughes:1987tj,Kuhl:1999kd} for details).

\begin{table}
\caption{\label{parameters}Values of the coefficients
$(\alpha^{}_m,\alpha^{}_f,\beta,\gamma)$ that characterize the different
evolution algorithms, in order to achieve consistency, stability, and
favorable high-frequency mode damping properties.}
\begin{ruledtabular}
\begin{tabular}{ccccc}
Algorithm & $\alpha^{}_m$ & $\alpha^{}_f$ & $\beta$ & $\gamma$ \\
\hline
Newmark & $0$ & $0$ & $(1+\rho^{}_\infty)^2$ & $\frac{3-\rho^{}_\infty}{2(1+\rho^{}_\infty)}$ \\
Bossak-$\alpha$ & $\frac{\rho^{}_\infty-1}{\rho^{}_\infty+1}$ & $0$
       & $\frac{1}{4}(1-\alpha^{}_m)^2$ & $\frac{1}{2}-\alpha^{}_m$ \\
Hilber-$\alpha$ & $0$ & $\frac{1-\rho^{}_\infty}{1+\rho^{}_\infty}$ &
       $\frac{1}{4}(1+\alpha^{}_f)^2$ & $\frac{1}{2}+\alpha^{}_f$ \\
Generalized-$\alpha$ & $\frac{2\rho^{}_\infty-1}{\rho^{}_\infty+1}$ &
       $\frac{\rho^{}_\infty}{1+\rho^{}_\infty}$ &
       $\frac{1}{4}(1-\alpha^{}_m+\alpha^{}_f)^2$ &
       $\frac{1}{2}-\alpha^{}_m+\alpha^{}_f$
\end{tabular}
\end{ruledtabular}
\end{table}

The numerical method we use to integrate the ODEs for the motion of the
particle, equations (\ref{ode1},\ref{ode2}), is the Bulirsch-Stoer
extrapolation method (\cite{Bulirsch:1966bs,Stoer:1993sb})
as described by~\cite{Ito:1997aj} and~\cite{Fukushima:1996aj} (see
also~\cite{Press:1992nr}).

\subsection{Structure and Motion of the Mesh}\label{meshstruct}
In the numerical simulations we have carried out we have used different mesh structures,
all motivated by the fact that the size of the particle is very small compared to
the length scale of the black hole.
The features that distinguish these different mesh structures are the following:
(i) {\em Refinement}.  Whether the size of the elements changes along the mesh in
order to increase the resolution around the particle.  (ii) {\em Particle's location}.
Whether the particle is located at the position of a node or, in contrast, it is
located in the interior of an element.  (iii) {\em Motion of the Mesh}.  Whether
the mesh is changing in time (in such a way that the part of the mesh with more resolution
always contains the particle) or it is static.

\begin{figure}[htb]
\begin{center}
\includegraphics[width=85mm]{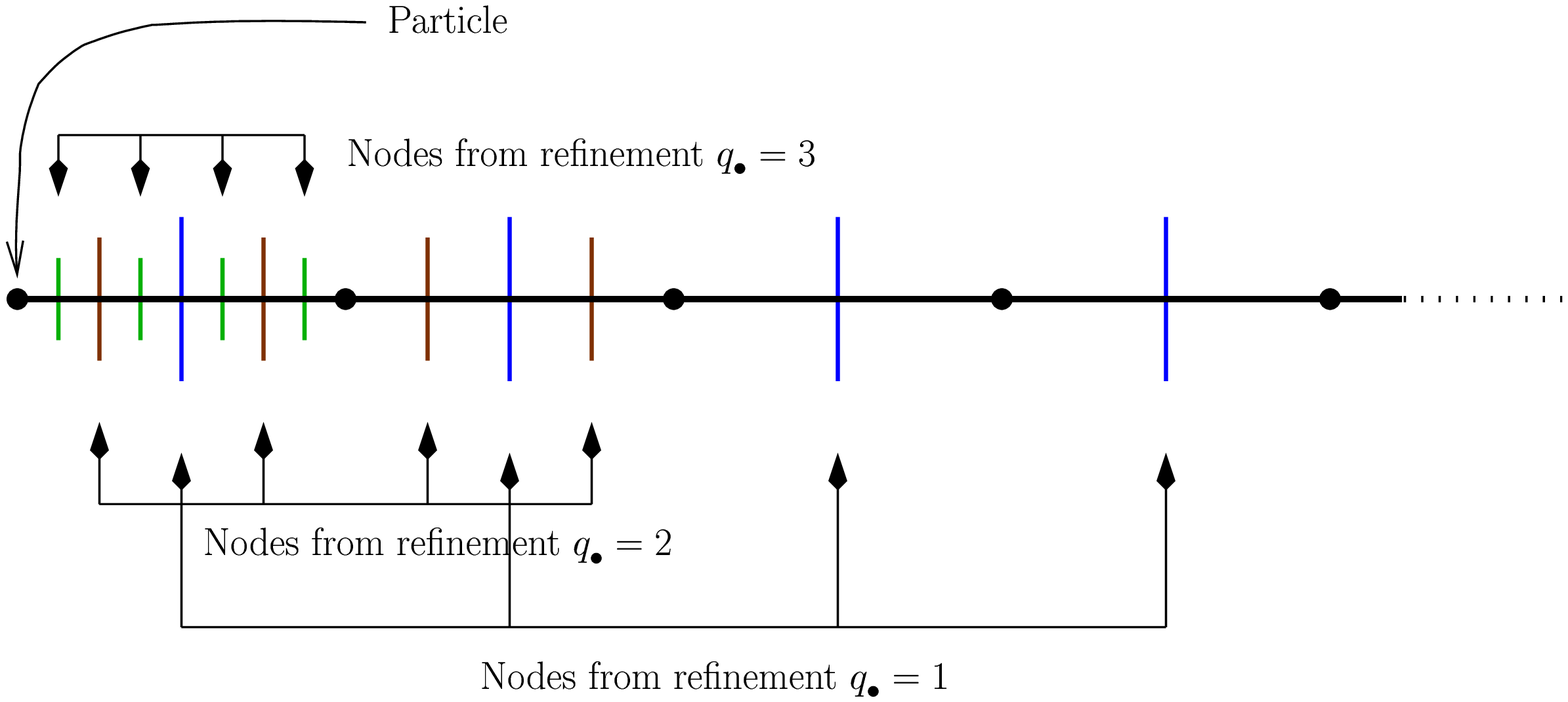}\\[5mm]
\includegraphics[width=85mm]{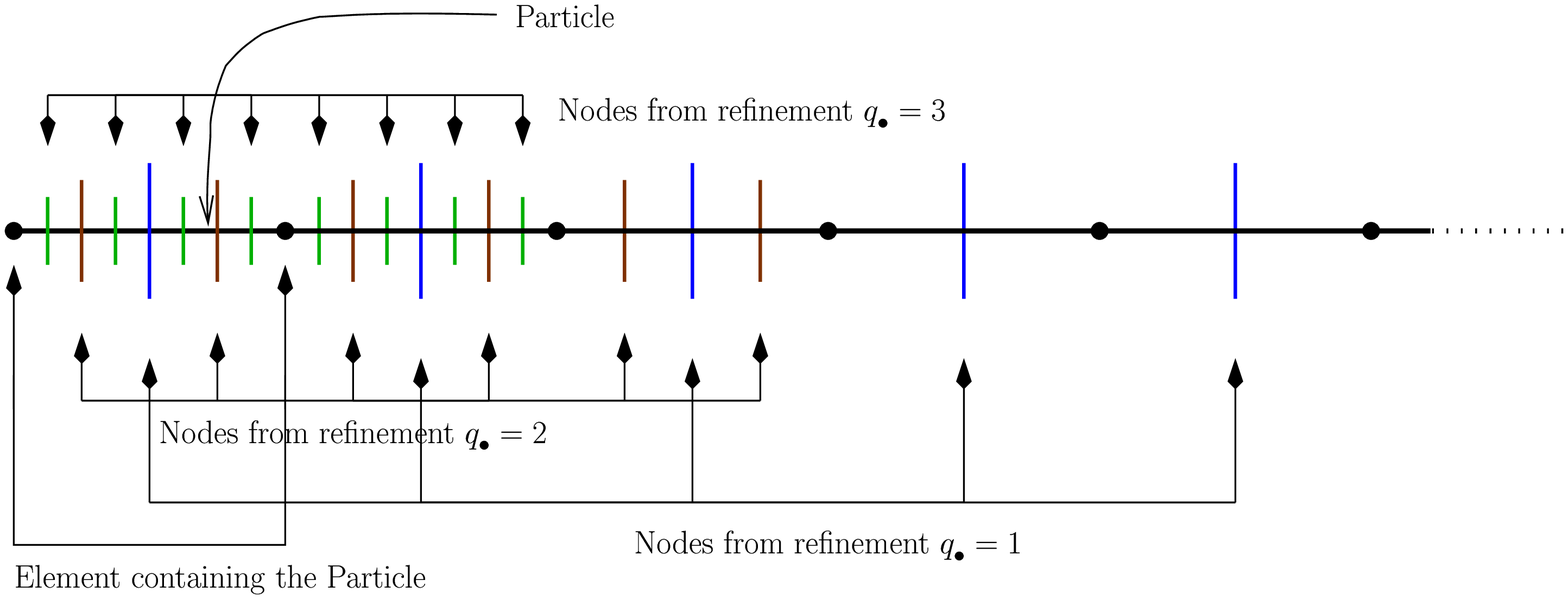}
\caption{Examples showing the structure of the Mesh for $(p^{}_\bullet,q^{}_\bullet)
= (4,3)$: On the top we have the case of a Mesh where the particle is located at
a node.  On the bottom we have the case of a Mesh where the particle is always in
the interior of an element. \label{mesh_particle}}
\end{center}
\end{figure}

In the case without refinement, we just divide the mesh into a given number
of elements, say $N^{}_o\,,$ with the same size: $d^{}_i = d\,,$ for all $i$.
In the case where the particle is located at a node, since the location of the
boundaries is also given, at least one element must have a size different
from $d$.  The way in which we refine the mesh to increase the resolution around
the particle is by dividing a certain number of elements  in the proximity of the
particle a certain number of times.   Each time, we divide each of the elements selected
into two elements of equal size.  For the case in which the particle is at a node, we
divide into two a given number of elements, say $p^{}_{\bullet}$, to the right and to
the left of the particle.  We repeat this a given number of times, say
$q^{}_{\bullet}\,.$  When the particle is located in the interior of a given element,
we do the same but with the elements to the right and to the left of the element where
the particle is located.  In addition, we bisect the element where the particle
is located $q^{}_\bullet$ times.   Then, the mesh is determined by the
three parameters $(N^{}_o,p^{}_\bullet,q^{}_\bullet)\,.$    The total number
of elements in the case where the particle is located at a node is:
$N^{}_T = N^{}_o + 2\,p^{}_\bullet\, q^{}_\bullet\,,$ and in the case where it is
located in the interior of an element is given by $N^{}_T = N^{}_o +
2\,p^{}_\bullet\, q^{}_\bullet + 2^{\,q^{}_\bullet}_{} - 1\,.$  We show an example of
these constructions in Figure~\ref{mesh_particle}.

Since the particle is moving, it may be convenient to adapt the mesh so that
the finest region is always around the particle.  We do this by just translating
the structure of the mesh but without modifying it, excepting for the elements
containing boundary nodes, whose size we need to change so that after the movement,
the mesh fits in our domain.  In other words,  the resulting mesh is the result
of applying a translation to all nodes and the translation distance is just the
distance the particle has moved.  After the mesh has been moved, we need to find
the new associated nodal functions by using the FE interpolation.
This is a very simple way of implementing a
moving mesh technique, but given that in this problem we know at every moment
where the resolution is required we do not need anything more sophisticated, at
least at this stage.

\section{Results from the Simulations}\label{results}
We have designed a numerical code with the ingredients described above.  To check
its performance we have carried out a number of different test.  First of all,
we have tested it with Gaussian profiles propagating in flat space  (both from
rest and with initial velocities), and also with Gaussian profiles scattering
off the potential of axial and polar modes.   To that end we used uniform meshes
and the following evolution schemes:  the average acceleration method, and the
Bossak-$\alpha$, Hilber-$\alpha$, and Generalized-$\alpha$ methods with different
values of $\rho^{}_\infty\,.$
In all these tests we found
stable and second-order convergent (in the $L^2$ norm) evolutions.
Deviations from second-order convergence were found to be of the order of
$0.1\%\,.$

When we introduce the point-like object we are introducing source terms that
contain a Dirac delta term and a term containing the first derivative of the
Dirac delta distribution.  These are singular terms, the first one induces a discontinuity
in the first derivative of the solution, whereas the second one induces a
discontinuity in the solution itself (see the description of these discontinuities
given above in Section~\ref{bhpert}).   This loss of smoothness of the solution
with respect to the case without particle is quite significant.  The presence
of Dirac delta distributions can still be handled by the FEM without losing
the convergence properties of the algorithms, but the inclusion of source
terms with the first derivative of the Dirac delta is too severe to maintain the
accuracy and convergence properties of the numerical algorithms
(see~\cite{StrFix:73} for a discussion of this issue).   As a consequence,
the convergence in general drops from second to first order.  There is however
a way of preserving second-order convergence, consisting of locating the
particle at a node and, instead of solving the equation with the {\em force}
term due to the particle, we solve the equation without the {\em force}
term in the region to the left and in the region to the right of the particle
location using boundary conditions at the particle location that enforce the
magnitude of the discontinuities of the solution that the particle source
terms dictate.   The way of computing these discontinuities is to use the
equations that govern their behaviour [equations~(\ref{discon1}) 
and~(\ref{discon2})].
However, this way of proceeding has some drawbacks depending on the way
we implement it.   Either it requires to change the structure of the matrices in
the FEM discretization of the equation, transforming the linear algebra problem
and making it similar to the one that we would get if we were using high-order
elements, or it changes the stability properties of our time-evolution schemes from
being unconditionally stable (they are implicit schemes) to be subject to a
Courant-Friedrichs-Levy stability condition.  Moreover, locating
the particle at a node means to change the mesh structure at every step
in the evolution (excepting for circular orbits), which means to use the FEM
interpolation every single time step.
The conclusion one can extract from this discussion is that each of the
different possible ways in which we can carry out the computations have some
advantages and some disvantages.   The performance of each of these possible
computational schemes depends on the physical setup we want to simulate and
therefore, one has analyze on a case by case basis which is the appropiate
method to use.

With regard to the choice of the time scheme, which in our framework is equivalent
to the choice of the parameters $(\beta,\gamma,\alpha^{}_m,\alpha^{}_f)$, the
numerical experiments we have performed tell us that the inclusion of the particle
generates high-frequency modes that corrupt the solution and therefore we have
to choose $\rho^{}_\infty$ different from unity to damp those unphysical modes.
In the case of the Newmark scheme, this means to loose second-order convergence,
and therefore it is not the best choice.   Moreover, from our numerical experiments
we observe that the Bossak-$\alpha$ scheme is the one that seems to work better in the
sense that it is the scheme that damps the high-frequency modes in a more efficient
way.  The other schemes seems to require a lower $\rho^{}_\infty$ (higher damping)
than the Bossak-$\alpha$ method for the same performance.   We have also seen
that the optimal value of $\rho^{}_\infty$ to be used depends on the physical
case we want to simulate, circular orbits are the ones that require less damping
whereas highly eccentric seems to require much more damping (for a comparable
pericenter distance).  It also depends on whether we move the mesh or not and on
the resolution we use.

In order to further assess the capabilities of our computational framework and its adequacy
for the type of physical computations we are interested in,
we have compared our numerical simulation with different results in the literature
for different types of orbits (geodesics).  The initial data for the master functions
is zero data, that is, $\Psi(t^{}_o,r) = \dot{\Psi}(t^{}_o,r) = 0\,.$  This creates
an initial burst of spurious radiation which, after suficient time, leaves the 
computational domain.  The global spatial resolution we have used in the simulation 
varies from $\Delta x = 0.1M$ to $\Delta x = 0.02M$, and the number of times that
we refine around the particle goes from $q^{}_\bullet = 0$ to $q^{}_\bullet = 10$
(see subsection~\ref{meshstruct}).
The physical observers or detectors of the gravitational radiation are located
at a tortoise coordinate in the range $|r^{}_\ast| =  2000-2500M\,,$ and the boundaries
are located at a distance in the range $|r^{}_\ast| = 4000-6000M\,.$
 Regarding the time
step, it is important to remark that because our evolution algorithms are implicit 
and unconditionally stable we are not subject to a Courant-Friedrich-Levy type condition 
on $\Delta t$ (excepting in the case where we use the scheme in which we impose the 
discontinuities generated by the particle at a given node), which we have taken to
be $\Delta t = 0.1M$.

To begin with, we compare results for circular orbits with the frequency domain code by
Poisson~\cite{Poisson:1995vs,Poisson:1997ad} (as quoted in~\cite{Martel:2003jj}), and with
the time-domain calculations by Martel~\cite{Martel:2003jj} (using a formulation based on
the Regge-Wheeler gauge and solving for the master functions) and Barack and
Lousto~\cite{Barack:2005nr} (using a formulation based on the Lorentz gauge and solving
directly for the metric perturbations).  The circular orbits considered have a radius $p M$ with
$p=7.9456$ and our observer is located in the radiation zone at $r^{}_{\!\ast}=2000M$.
We compare results for the energy and angular momentum luminosities to
infinity with the results of~\cite{Poisson:1995vs,Poisson:1997ad} and~\cite{Martel:2003jj}
in Table~\ref{circularorbits1}, and for the energy luminosities to infinity
with the results of~\cite{Barack:2005nr} in Table~\ref{circularorbits2}.

\begin{table*}
\caption{\label{circularorbits1}Comparison of the computations of energy and angular
momentum luminosities at infinity  for circular orbits with $p=7.9456$
with results obtained with the frequency-domain code by Poisson~\cite{Poisson:1995vs,Poisson:1997ad}
and the time-domain code by Martel~\cite{Martel:2003jj}.  They are calculated at
$r^{}_{\!\ast}=2000M$.  The energy luminosities are expressed in units of $(M/\mu)^2$
and the angular momentum luminosities in units of $M/\mu^2$.
In square brackets we have included the absolute relative difference
(rounded to the largest value).}
\begin{ruledtabular}
\begin{tabular}{c|c|c|c|c|c|c}
$(\ell,m)$~ & $\dot{E}^{\infty}_{\ell m}$ & $\dot{L}^{\infty}_{\ell m}$ &
$\dot{E}^{\infty}_{\ell m}$ \cite{Poisson:1995vs,Poisson:1997ad} &
$\dot{L}^{\infty}_{\ell m}$ \cite{Poisson:1995vs,Poisson:1997ad} &
$\dot{E}^{\infty}_{\ell m}$ \cite{Martel:2003jj} &
$\dot{L}^{\infty}_{\ell m}$ \cite{Martel:2003jj} \\
\hline
$(2,1)$   & $8.1662\cdot 10^{-7}$  & $1.8289\cdot 10^{-5}$  & $8.1633\cdot 10^{-7}$  [$0.04$\%]   & $1.8283\cdot 10^{-5}$  [$0.04$\%]  & $8.1623\cdot 10^{-7}$  [$0.05$\%] & $1.8270\cdot 10^{-5}$  [$0.1$\%] \\
$(2,2)$   & $1.7064\cdot 10^{-4}$  & $3.8219\cdot 10^{-3}$  & $1.7063\cdot 10^{-4}$  [$0.006$\%]  & $3.8215\cdot 10^{-3}$  [$0.01$\%]  & $1.7051\cdot 10^{-4}$  [$0.08$\%] & $3.8164\cdot 10^{-3}$  [$0.2$\%] \\
$(3,1)$   & $2.1732\cdot 10^{-9}$  & $4.8675\cdot 10^{-8}$  & $2.1731\cdot 10^{-9}$  [$0.005$\%]  & $4.8670\cdot 10^{-8}$  [$0.01$\%]  & $2.1741\cdot 10^{-9}$  [$0.05$\%] & $4.8684\cdot 10^{-8}$  [$0.02$\%] \\
$(3,2)$   & $2.5204\cdot 10^{-7}$  & $5.6450\cdot 10^{-6}$  & $2.5199\cdot 10^{-7}$  [$0.02$\%]   & $5.6439\cdot 10^{-6}$  [$0.02$\%]  & $2.5164\cdot 10^{-7}$  [$0.2$\%]  & $5.6262\cdot 10^{-6}$  [$0.4$\%]  \\
$(3,3)$   & $2.5475\cdot 10^{-5}$  & $5.7057\cdot 10^{-4}$  & $2.5471\cdot 10^{-5}$  [$0.02$\%]   & $5.7048\cdot 10^{-4}$  [$0.02$\%]  & $2.5432\cdot 10^{-5}$  [$0.2$\%]  & $5.6878\cdot 10^{-4}$  [$0.4$\%]  \\
$(4,1)$   & $8.4055\cdot 10^{-13}$ & $1.8825\cdot 10^{-11}$ & $8.3956\cdot 10^{-13}$ [$0.12$\%]   & $1.8803\cdot 10^{-11}$ [$0.12$\%]  & $8.3507\cdot 10^{-13}$ [$0.7$\%]  & $1.8692\cdot 10^{-11}$ [$0.7$\%]  \\
$(4,2)$   & $2.5099\cdot 10^{-9}$  & $5.6215\cdot 10^{-8}$  & $2.5091\cdot 10^{-9}$  [$0.04$\%]   & $5.6195\cdot 10^{-8}$  [$0.04$\%]  & $2.4986\cdot 10^{-9}$  [$0.5$\%]  & $5.5926\cdot 10^{-8}$  [$0.6$\%]  \\
$(4,3)$   & $5.7765\cdot 10^{-8}$  & $1.2937\cdot 10^{-6}$  & $5.7751\cdot 10^{-8}$  [$0.03$\%]   & $1.2934\cdot 10^{-6}$  [$0.03$\%]  & $5.7464\cdot 10^{-8}$  [$0.6$\%]  & $1.2933\cdot 10^{-6}$  [$0.03$\%] \\
$(4,4)$   & $4.7270\cdot 10^{-6}$  & $1.0586\cdot 10^{-4}$  & $4.7256\cdot 10^{-6}$  [$0.03$\%]   & $1.0584\cdot 10^{-4}$  [$0.02$\%]  & $4.7080\cdot 10^{-6}$  [$0.4$\%]  & $1.0518\cdot 10^{-4}$  [$0.7$\%]  \\
$(5,1)$   & $1.2607\cdot 10^{-15}$ & $2.8237\cdot 10^{-14}$ & $1.2594\cdot 10^{-15}$ [$0.1$\%]    & $2.8206\cdot 10^{-14}$ [$0.1$\%]   & $1.2544\cdot 10^{-15}$ [$0.5$\%]  & $2.8090\cdot 10^{-14}$ [$0.6$\%] \\
$(5,2)$   & $2.7909\cdot 10^{-12}$ & $6.2509\cdot 10^{-11}$ & $2.7896\cdot 10^{-12}$ [$0.05$\%]   & $6.2479\cdot 10^{-11}$ [$0.05$\%]  & $2.7587\cdot 10^{-12}$ [$1.2$\%]  & $6.1679\cdot 10^{-11}$ [$1.4$\%] \\
$(5,3)$   & $1.0936\cdot 10^{-9}$  & $2.4494\cdot 10^{-8}$  & $1.0933\cdot 10^{-9}$  [$0.03$\%]   & $2.4486\cdot 10^{-8}$  [$0.04$\%]  & $1.0830\cdot 10^{-9}$  [$1.0$\%]  & $2.4227\cdot 10^{-8}$  [$1.0$\%] \\
$(5,4)$   & $1.2329\cdot 10^{-8}$  & $2.7613\cdot 10^{-7}$  & $1.2324\cdot 10^{-8}$  [$0.04$\%]   & $2.7603\cdot 10^{-7}$  [$0.04$\%]  & $1.2193\cdot 10^{-8}$  [$1.1$\%]  & $2.7114\cdot 10^{-7}$  [$1.8$\%] \\
$(5,5)$   & $9.4616\cdot 10^{-7}$  & $2.1190\cdot 10^{-5}$  & $9.4563\cdot 10^{-7}$  [$0.06$\%]   & $2.1179\cdot 10^{-5}$  [$0.06$\%]  & $9.3835\cdot 10^{-7}$  [$0.9$\%]  & $2.0933\cdot 10^{-5}$  [$1.3$\%] \\
\hline
Total     & $2.0293\cdot 10^{-4}$  & $4.5451\cdot 10^{-3}$  & $2.0292\cdot 10^{-4}$ [$0.005$\%]   & $4.5446\cdot 10^{-3}$  [$0.02$\%]  & $2.0273\cdot 10^{-4}$  [$0.1$\%]  & $4.5399\cdot 10^{-3}$  [$0.2$\%]
\end{tabular}
\end{ruledtabular}
\end{table*}

\begin{table}
\caption{\label{circularorbits2}Comparison of the computations of energy luminosities
[expressed in units of $(M/\mu)^2$] at infinity for circular orbits with $p=7.9456$
with results obtained in the time domain by Barack and Lousto~\cite{Barack:2005nr}.
They are calculated at $r^{}_{\!\ast}=2000M$.  In square brackets we have included the
absolute relative difference (rounded to the largest value).}
\begin{ruledtabular}
\begin{tabular}{c|c|c}
$(\ell,m)$~ & $\dot{E}^{\infty}_{\ell m}$ &
$\dot{E}^{\infty}_{\ell m}$ \cite{Barack:2005nr} \\
\hline
$(2,1)$   & $8.1662\cdot 10^{-7}$  & $8.1654\cdot 10^{-7}$ [$0.01$\%] \\
$(2,2)$   & $1.7064\cdot 10^{-4}$  & $1.7061\cdot 10^{-4}$ [$0.02$\%] \\
$(3,1)$   & $2.1732\cdot 10^{-9}$  & $2.1734\cdot 10^{-9}$ [$0.01$\%] \\
$(3,2)$   & $2.5204\cdot 10^{-7}$  & $2.5207\cdot 10^{-7}$ [$0.02$\%] \\
$(3,3)$   & $2.5475\cdot 10^{-5}$  & $2.5479\cdot 10^{-5}$ [$0.02$\%] \\
$(4,1)$   & $8.4055\cdot 10^{-13}$ & $8.3982\cdot 10^{-13}$ [$0.09$\%] \\
$(4,2)$   & $2.5099\cdot 10^{-9}$  & $2.5099\cdot 10^{-9}$ [$0.004$\%] \\
$(4,3)$   & $5.7765\cdot 10^{-8}$  & $5.7759\cdot 10^{-8}$ [$0.01$\%]  \\
$(4,4)$   & $4.7270\cdot 10^{-6}$  & $4.7284\cdot 10^{-6}$ [$0.03$\%]  \\
$(5,1)$   & $1.2607\cdot 10^{-15}$ & $1.2598\cdot 10^{-15}$ [$0.07$\%] \\
$(5,2)$   & $2.7909\cdot 10^{-12}$ & $2.7877\cdot 10^{-12}$ [$0.12$\%]   \\
$(5,3)$   & $1.0936\cdot 10^{-9}$  & $1.0934\cdot 10^{-9}$ [$0.02$\%] \\
$(5,4)$   & $1.2329\cdot 10^{-8}$  & $1.2319\cdot 10^{-8}$ [$0.08$\%] \\
$(5,5)$   & $9.4616\cdot 10^{-7}$  & $9.4623\cdot 10^{-7}$ [$0.008$\%] \\
\hline
Total     & $2.0293\cdot 10^{-4}$  & $2.0291\cdot 10^{-4}$ [$0.01$\%]
\end{tabular}
\end{ruledtabular}
\end{table}

We have also compared results for elliptic orbits with the frequency-domain calculations
of Cutler et al.~\cite{Cutler:1994pb} and with the time-domain calculations
of Martel~\cite{Martel:2003jj}.   We have considered two types of elliptic orbits
with orbital parameters given by $(p,e) = (7.50478,0.188917)$ and
$(p,e) = (8.75455,0.764124)\,.$  For these orbits we have computed the averaged
energy and angular momentum luminosities.  The average is taken over a certain number
of radial periods and our observer is located at $r^{}_{\!\ast}=2500M$.
The results are shown in Table~\ref{ellipticorbits}.

\begin{table*}
\caption{\label{ellipticorbits}Computations of the total
{\em average} energy [in units of $(M/\mu)^2$] and angular momentum luminosities
[in units of $M/\mu^2$], $<\dot{E}^{\infty}>$ and $<\dot{L}^{\infty}>$,
for elliptic orbits.  They are calculated at $r^{}_{\!\ast}=2500M$. We compare with the
results obtained by Cutler et al.~\cite{Cutler:1994pb} using
a frequency-domain numerical code and by Martel~\cite{Martel:2003jj} using a time-domain
numerical code.  We consider two different types of elliptic orbits: Orbit $A\,$:
$(p,e) = (7.50478,0.188917)\,.$  Orbit $B\,$:  $(p,e) = (8.75455,0.764124)$.}
\begin{ruledtabular}
\begin{tabular}{c|c|c|c|c|c|c}
Orbit~ & $<\dot{E}^{\infty}>$ &  $<\dot{L}^{\infty}>$ &
$<\dot{E}^{\infty}>$ \cite{Cutler:1994pb} &
$<\dot{L}^{\infty}>$ \cite{Cutler:1994pb} &
$<\dot{E}^{\infty}>$ \cite{Martel:2003jj} &
$<\dot{L}^{\infty}>$ \cite{Martel:2003jj} \\
\hline
$A$ & $3.1640\cdot 10^{-4}$  & $5.9555\cdot 10^{-3}$  & $3.1680\cdot 10^{-4}$  [$0.2$\%] & $5.9656\cdot 10^{-3}$ [$0.2$\%] & $3.1770\cdot 10^{-4}$ [$0.5$\%] & $5.9329\cdot 10^{-3}$ [$0.4$\%]\\
$B$ & $2.1004\cdot 10^{-4}$  & $2.7505\cdot 10^{-3}$  & $2.1008\cdot 10^{-4}$  [$0.02$\%] & $2.7503\cdot 10^{-3}$ [$0.01$\%] & $2.1484\cdot 10^{-4}$ [$2.3$\%] & $2.7932\cdot 10^{-3}$ [$1.6$\%]
\end{tabular}
\end{ruledtabular}
\end{table*}

\begin{figure}[htb]
\begin{center}
\includegraphics[width=85mm]{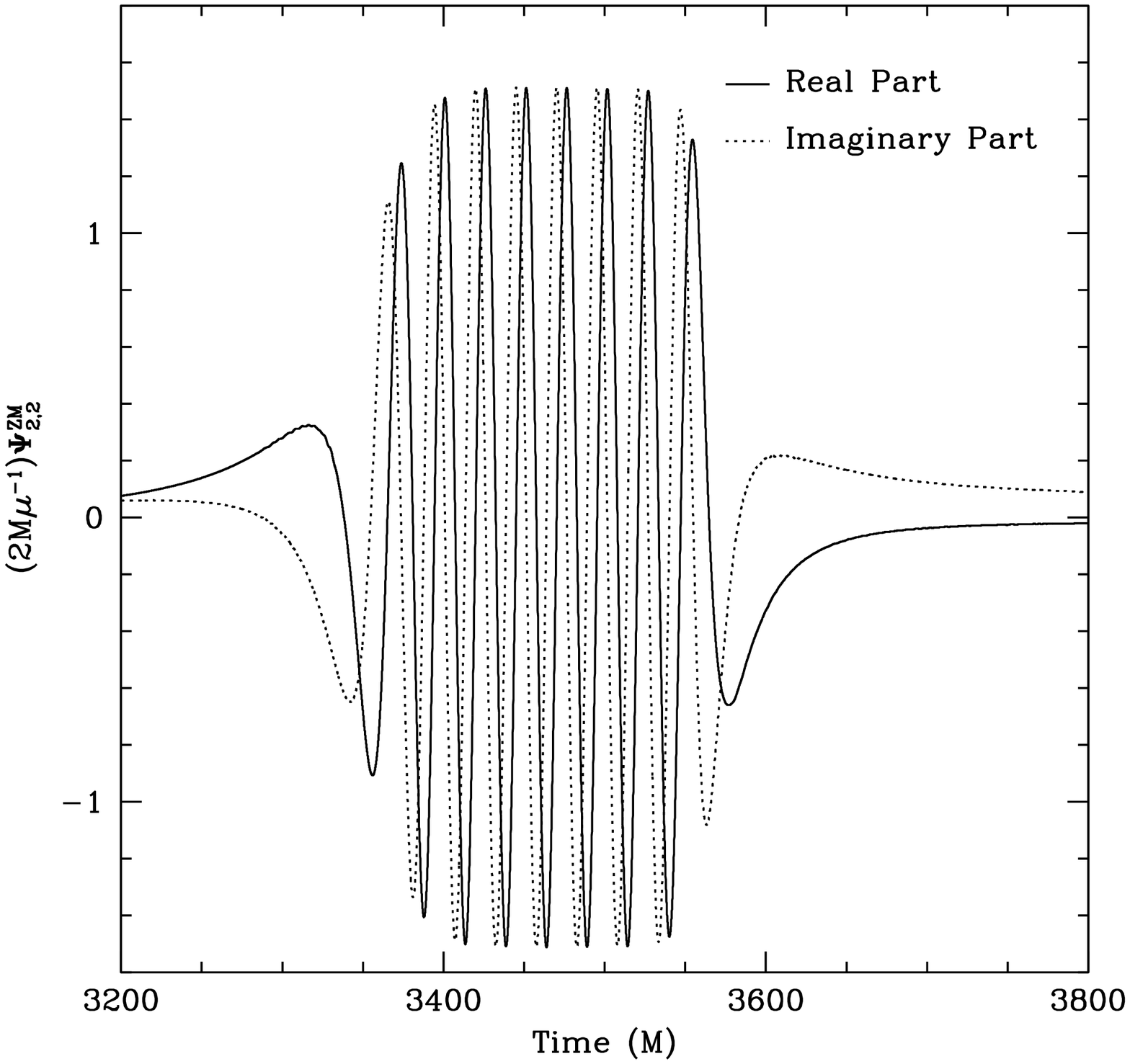}
\caption{Component $(\ell,m)=(2,2)$ of the waveforms corresponding to {\em zoom-whirl}
parabolic orbits ($e=1$) with $p=8.001$. \label{waveform_zoomwhirl}}
\end{center}
\end{figure}

We also compare results for parabolic orbits with the time-domain calculations
of Martel~\cite{Martel:2003jj}.  This type of orbits have $e=1$ and are only
characterized by the pericenter distance, which is given by $p M/2$ with $p>8$.
As $p$ approaches $8$ the number of orbital periods ($\Delta\varphi^{}_p/2\pi$)
diverges and the motion shows the so-called {\em zoom-whirl} behaviour
(see, e.g.~\cite{Martel:2003jj}), meaning that for a radial period the particle
orbits close to the MBH for a number of orbital periods producing a very characteristic signal
(see Figure~\ref{waveform_zoomwhirl})
with a number of cycles that depends on how close $p$ is to $8$.
They are therefore a good test bed for the numerical computations.
In Table~\ref{parabolicorbits} we show the computations of the total
energy and angular momentum radiated to infinity $(E^{\infty},L^{\infty})$
and into the horizon $(E^{H},L^{H})$ for parabolic orbits with
$p=8.00001$ and $p=8.001\,.$   For these computations, our observers are located at
$r^{}_{\!\ast}=-2500M$ and $r^{}_{\!\ast}=2500M$.

\begin{table*}
\caption{\label{parabolicorbits}Computations of the total
energy [in units of $M/\mu^2$] and angular momentum [in units of $\mu^{-2}$] radiated,
both to infinity $(E^{\infty},L^{\infty})$ and into the horizon $(E^{H},L^{H})$,
in parabolic orbits ($e=1$). They are calculated at $r^{}_{\!\ast}=-2500M$
and $r^{}_{\!\ast}=2500M$.  In square brackets we have included the absolute
relative difference (rounded to the largest value) with respect the results
obtained by Martel~\cite{Martel:2003jj} using a time-domain numerical code.}
\begin{ruledtabular}
\begin{tabular}{c|c|c|c|c}
$p$~      & $E^{\infty}$       & $L^{\infty}$       & $E^{H}$                          & $L^{H}$                         \\
\hline
$8.00001$ & $3.5603$ [$3.1$\%] & $29.415$ [$2.5$\%] & $1.8884\cdot 10^{-1}$ [$0.05$\%] & $1.5112$ [$0.7$\%]              \\
$8.001$   & $2.2212$ [$2.7$\%] & $18.704$ [$2.1$\%] & $1.1339\cdot 10^{-1}$ [$0.7$\%]  & $9.0783\cdot 10^{-1}$ [$0.5$\%]
\end{tabular}
\end{ruledtabular}
\end{table*}

\begin{figure}[htb]
\begin{center}
\includegraphics[width=85mm]{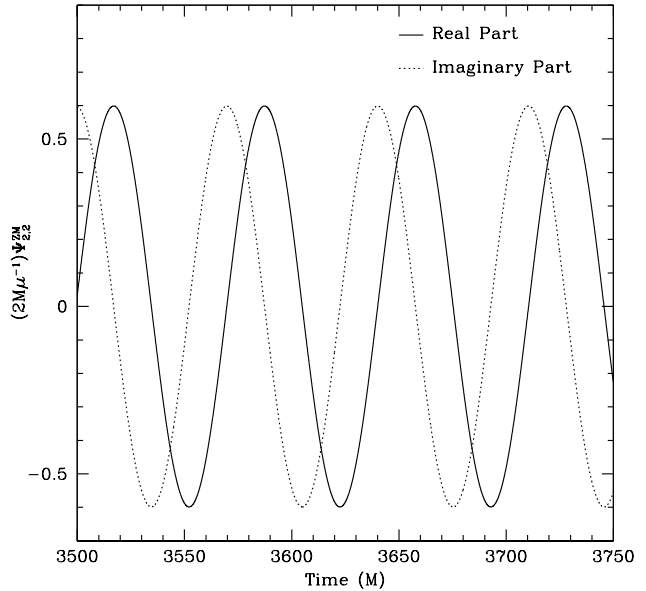}
\caption{Component $(\ell,m)=(2,2)$ of the waveform corresponding to circular orbits 
($e=0$) with $p=7.9456\,$. \label{waveform_circular}}
\end{center}
\end{figure}

\begin{figure}[htb]
\begin{center}
\includegraphics[width=85mm]{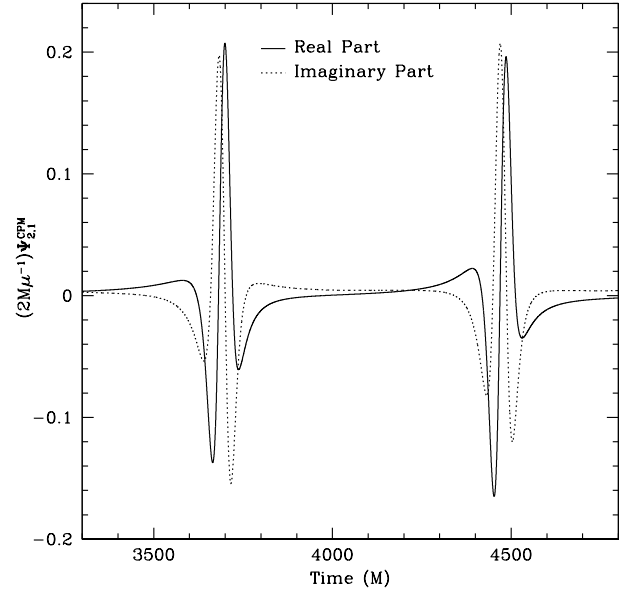}
\caption{Component $(\ell,m)=(2,1)$ of the waveform corresponding to elliptic orbits 
with $e=0.764124$ and $p=8.75455\,$. \label{waveform_elliptic}}
\end{center}
\end{figure}

We finish this section by commenting on the waveforms obtained
from our numerical computations.  We have already mentioned that one of the advantages
of the time-domain approach is that it can provide reliable waveforms for a reasonable
computational cost.  We show that this is indeed the case by plotting the following
components of the waveforms: $\Psi^{\ZM}_{2,2}$ for
circular orbits with $p=7.9456\,$,  $\Psi^{\CPM}_{2,1}$ for elliptic orbits with 
$(e,p) = (0.764124,8.75455)\,$, and $\Psi^{\ZM}_{2,2}$ for parabolic orbits
with $p=8.001$ in Figures~\ref{waveform_circular},~\ref{waveform_elliptic},
and~\ref{waveform_zoomwhirl} respectively.  To achieve a high degree of smoothness
in the waveforms, the damping of the spourious high-frequency modes in the
evolution is crucial.  In this sense, our simulations show that the evolution
numerical algorithms proposed in this paper are suitable for the production of
reliable waveforms.

\section{Conclusions and Discussion}\label{discussion}
In this paper we have presented a new method for computing the gravitational radiation
emitted by a point like object orbiting a non-rotating black hole.  We have shown
that the method is accurate by comparing it with previous results in the literature
obtaining an agreement with relative errors of the order of $1$\%, in many
cases even of the order of $0.1$\% or below.  We also have shown that these numerical techniques
provide sufficiently smooth waveforms, which is one of the goals of these calculation
in relation with gravitational-wave data analysis efforts.
These results together with the particular feature of the computational method presented
suggest that it is a suitable method to be use in self-force
calculations for inspiralling EMRBs and in the posterior waveform calculations at
the next perturbative order.
Our numerical calculations are based on the FEM and related techniques.
The main features of the FEM that makes it suitable for the study of EMRBs, and perhaps
also for problems that Numerical Relativity deals with, are the following:
(i) {\em Proper description of the Computational Domain}.  This is particularly
relevant when we want to solve the perturbative equations in a 2D or 3D setup
(see~\cite{Sopuerta:2005rd}),
as it is the case if we want to consider a rotating Kerr black hole, which is the
astrophysically relevant case.  It would be also relevant for the study of black
hole spacetimes in Numerical Relativity.  In this scenario, the spacetime geometry may
involve holes (inner boundaries arising from black hole singularity excision) and
we may wish to use a spherical-type outer boundary to allow gravitational radiation
leave the domain smoothly.    All these geometric issues have usually caused a number
of problems in Finite Differences techniques, but can be handled in a natural
way using the FEM.   In this respect, the FEM has already shown its capabilities
in solving problems in other scientific areas that involve much more complicated domains
than the ones we can face in General Relativity.
(ii) {\em Imposition of Boundary Conditions}. In close connection with the previous
point, the underlaying philosophy in the FEM is that one should use
the mesh that adapts best to the geometric characteristics of the problem we want
to solve.  In  particular, to the boundary conditions, since it is not equally
simple and convenient to impose outgoing radiation conditions in a rectangular boundary
than in a spherical one.  This also has an impact when we perform the FEM discretization,
since it is based on the {\em weak} form of our equations, which can have built in the
the boundary conditions.  In the case of problems in 2D or higher dimensions, if the boundary
is {\em natural} (adapted to the problem), the implementation
of the boundary conditions becomes trivial (see, e.g.~\cite{Sopuerta:2005rd}).
This has advantages even in 1D problems, like the one we have studied in this
paper, where the imposition of boundary conditions like Sommerfeld or von Neumann
is simpler than in a Finite Differences framework.  This paper illustrates this fact.
(iii) {\em Treatment of distributions.}  Many description of EMRBs treat the small body
as a point-like object, which despite being somehow unnatural in General Relativity,
allows us to perform computations in a consistent way.   The consequence of having a point-like
object is that the equations that we have to solve contain source terms where Dirac delta
distributions and its derivatives (up to second derivatives in the case we were solving the
Teukolsky equation sourced by a point-like object) appear.   To deal with this kind of
distributions in a Finite Difference framework is not an easy task, and the different ways
in which one can handle them involved not trivial {\em a priori} regularizations of the
distributions.
Instead, in the FEM, the fact that the discretization is based on the {\em weak} form
of the equations, an integral formulation, is a key point.  We can evaluate the integrals
that involve Dirac distributions analytically by using the properties of the distributions,
without the need of using any regularization of those distributions.  A sample of this
has been given in this paper, where we used the {\em weak} formulation of the problem
to discretize a source term containing the Dirac delta distribution and its first
derivative.   Then, the type of discretization we would get is in this sense analogous
to the one proposed by Price and Lousto~\cite{Lousto:1997wf,Martel:2001yf,Martel:2003jj},
where they also used an integral form of the equations to discretize them.
Therefore, using the FEM provides an additional advantage for the study of EMRBs where
the small object is treated as a point-like object.
(iv) {\em Adaptivity.}   This is a key ingredient for the simulations of EMRBs.  The
calculations presented in this paper do not necessarily require adaptivity, but they
are an excellent benchmark to test these techniques.   However, for the case of rotating
massive black hole adaptivity may be the only way of performing physically realistic
simulations.  The FEM is a natural choice to achieve the high level of adaptivity
required, both in the construction of the mesh and later by using any of the robust
techniques of mesh refinement available (see~\cite{Sopuerta:2005rd} and references
therein).

Apart from these specific reasons, there are other motivations in favour of using
the FEM.   In this sense it is important to mention that because the FEM is based on
piecewise (polynomial) approximations, it lies on sound mathematical grounds
(see, e.g.~\cite{StrFix:73,Ciarlet:1978bo}).
From the point of view of building numerical codes based on the FEM, it is important
to emphasize the high degree of independence of the different ingredients of the
FEM discretization process~\cite{StrFix:73,Hughes:1987tj,Babuska:2001bs,Zienkiewicz:77oc},
which makes it very suitable for modular programming.  In addition, the FEM has been
widely used in many areas of scientific research and, as a consequence, a number of
FEM packages and tools are available for scientific computation.

There are a number of ways of extending this work in order to improve the computational
framework in order to simulate EMRBs, and in particular to evalute the radiation-reaction
effects.  From the computational side we can introduce higher-order elements (by using
FE functional spaces with higher-order polynomials), which will improve the accuracy of the computations.
From the physical point of view, we can change the description of the gravitational
field, meaning the formulation of the perturbative scheme.  In this sense, to compute
the metric perturbations using the Lorentz gauge, as it has been recently proposed
by Barack and Lousto~\cite{Barack:2005nr}, appear to be a very convenient choice for
a number of reasons (see~\cite{Barack:2005nr} for a detailed discussion).  Among the
advantages of this approach it is worth to mention the following ones: (i)  Because
one is working with pure metric perturbations the sources do not contain derivatives
of the Dirac delta distribution, and hence the solution of the equations is continuous
at the particle location, which will improve the accuracy of the computations.
(ii) Moreover, in contrast with the computations in the Regge-Wheeler gauge, we do not need
a metric perturbation reconstruction procedure (just algebraic computations) to evaluate
the self-force. (iii) The regularization procedures to obtain the self-force have only been
given in the Lorentz gauge.   It also has some disavantages:  We need to solve a coupled
system of equations instead of single wave-type equations, and there are constraints that
need to be satisfied along the evolution.

In the astrophysically motived EMRBs, the MBH is highly rotating and therefore it is
desirable to be able to repeat these calculations by using the Kerr solution as the
background spacetime.  This is a more difficult problem since it involves three-dimensional
PDEs (or two-dimensional if we factor out the dependence in the polar angle).
In this sense, it is important to mention that the FEM techniques
that have been presented and used in this paper can be transfered to the higher-dimensional
problem of computing Kerr perturbations.  For the same reasons that have been pointed
out before, a promising approach may be to solve for metric perturbations of the
Kerr black hole in the Lorentz gauge.

\[ \]
{\bf Acknowledgements:}
The authors acknowledge the support of the Center for Gravitational Wave Physics
funded by the National Science Foundation under Cooperative Agreement
PHY-0114375, and support from NSF grant PHY-0244788 to Penn State University.
They also wish to thank the Information Technology Services at Penn State University
for the use of the LION-XO computer cluster in some of the calculations presented
in this paper, and to Uli Sperhake, Pengtao Sun, and Jinchao Xu for fruitful
discussions.

\appendix

\section{Motion of the Point-like Object}\label{geodesic}
To complete the description of our physical problem we have to introduce the 
equations of motion for the point-like object, which follows the geodesics 
of the Schwarzschild background black hole spacetime.  Then, the four-velocity of the
particle satisfies:
\begin{equation}
u^\alpha\nabla_\alpha u^\beta = 0\,,~~~
u^\alpha = \frac{dz^\alpha(\tau)}{d\tau}\,.
\end{equation}
The static and spherically symmetric character of the background imply the
existence of first integrals of the motion (energy and angular momentum),
and as it happens in the Newtonian case, the motion takes place on a plane 
that, without lost of generality, we can take it to be the plane $\theta=\pi/2$ 
($u^\theta = 0$).  Then, the equations of motion are equivalent to the following
relations: 
\begin{equation}
u^t = \frac{E^{}_p}{f}\,,~
u^\varphi = \frac{L^{}_p}{r^2}\,,~
(u^r)^2 = E^{2}_p - f\left(1+\frac{L^2_p}{r^2}\right)\,. \label{eqmotion}
\end{equation}
In order to obtain a well-behaved system of ODEs at the turning points 
of the radial coordinate ($\dot{r}=0$) we can use the following alternative quantity
\begin{equation}
r = \frac{pM}{1+e\cos\chi}\,,
\end{equation}
where $e$ denotes the orbital eccentricity and $p$ the {\em semi-latus rectum},
which can be used as alternative constants of motion to the pair $(E^{}_p,L^{}_p)$.
Then, the two equations we need to integrate to determine the position of the
particle are:
\begin{equation}
\frac{d\chi}{dt} = \frac{(p-2-2e\cos\chi)(1+e\cos\chi)^2
\sqrt{p-6-2e\cos\chi}}{Mp^2\sqrt{(p-2)^2-4e^2}}
\,, \label{ode1}
\end{equation}
\begin{equation}
\frac{d\varphi}{dt} = \frac{(p-2-2e\cos\chi)(1+e\cos\chi)^2}
{Mp^{3/2}\sqrt{(p-2)^2-4e^2}} \,. \label{ode2}
\end{equation}

\section{Gauss-Legendre Quadratures} \label{GaussLegendre}
The integrals in the second terms of (\ref{stiffness}) are computed by using
Gauss-Legendre quadratures (see, e.g.~\cite{Hughes:1987tj} and~\cite{Press:1992nr}).
Given a function $F(x)$ we approximate its integral over the interval $[a,b]$
by
\begin{equation}
\int^b_a \hspace{-2mm} dx F(x) \approx \frac{b-a}{2}\sum_{I=1}^{N} W^{N}_I
F\left(\frac{a+b}{2} + \frac{b-a}{2}u^{}_I\right)\,,
\end{equation}
where $u^{}_I$ is the $I$-th zero of the Legendre polynomial $P^{}_N(u)$ (it has
exactly $N$ zeros) and $W^{N}_I$ are weights associated with the zeros and given by
\begin{equation}
W^N_I = \frac{2}{(1-u^2_I)\left[P'_N(u^{}_I)\right]^2} \,.
\end{equation}
An N-point Gauss-Legendre quadrature integrates exactly polynomials of degree
$2N-1\,.$


\end{document}